\documentclass[a4paper,10pt]{article}

\usepackage[onehalfspacing]{setspace}
\usepackage{amsmath}
\usepackage{amssymb}
\usepackage{braket}
\usepackage[T1]{fontenc}
\usepackage[utf8]{inputenc}
\usepackage{authblk}
\usepackage{color}
\usepackage{xcolor}
\usepackage{graphicx}
\usepackage{cite}
\usepackage{slashed}
\usepackage{tikz}
\usepackage{multirow}
\usepackage{hhline}
\usepackage[export]{adjustbox}
\usepackage[top=1.5cm,bottom=2.5cm,right=2.54cm,left=2.54cm]{geometry}
\usepackage[colorlinks=true,linkcolor=blue,citecolor=blue,urlcolor=blue]{hyperref}
\usepackage{tocloft}

\date{}

\title{Investigation of quantum chaos in local and non-local Ising models}  
\author[1,2]{Reza Pirmoradian } 
\author[2]{Elham Sadoogh}  
\author[2]{Maryam Teymouri} 
\author[2]{Negar Abolqasemi-Azad}
\author[2]{Mohammad Reza Lahooti}  
\author[2]{Zahra Mohammad-Ali}
\affil[1]{School of Quantum Physics and Matter Institute for Research in Fundamental Sciences (IPM), P.O. Box 19395-5531, Tehran, Iran}
\affil[2]{Faculty of engineering, Ershad Damavand Institute of Higher Education, Tehran, Iran,  P.O. Box 14676-86831, Tehran, Iran}

\begin{document}
\maketitle
\begin{center}
E-mails: {\tt rezapirmoradian@ipm.ir, sadooghelham@gmail.com, mm.teymouri@gmail.com, n.abolghasemiazad@gmail.com,drlahooti\_2003@yahoo.com, mohamadali.zha75@gmail.com}
\end{center}

\begin{abstract}
We investigate signatures of quantum chaos within Ising spin chains subjected to transverse and longitudinal fields, incorporating both local (nearest-neighbor) and non-local (long-range) couplings. While local Ising models may exhibit integrable or chaotic dynamics contingent on interaction strengths and field parameters, systems with non-local interactions generally display a stronger propensity toward chaos, even when the non-local couplings are weak. By examining the distribution of energy level spacings through the level spacing ratio, we delineate the transition from integrable to chaotic regimes and characterize the emergence of quantum chaos in these systems. Our analysis demonstrates that non-local couplings facilitate faster operator spreading and more intricate dynamical behavior, enabling these systems to approach maximal chaos more readily than their local counterparts. Additionally, we analyze Krylov complexity as a dynamical probe of chaos, observing a characteristic peak followed by a plateau at late times in chaotic regimes. This behavior provides a quantitative means to distinguish between integrable and chaotic phases, with the growth rate and saturation level of the complexity serving as effective indicators. Our findings underscore the role of non-local interactions in accelerating the onset of chaos and modifying dynamical complexity in quantum spin chains.
\end{abstract}
\newpage

\noindent\rule{\textwidth}{0.6pt}
\vspace{0.5cm}

\tableofcontents

\vspace{0.5cm}
\noindent\rule{\textwidth}{0.6pt}

\section{Introduction}
According to daily experience, the thermalization of macroscopic systems is the most natural phenomena. To observe a macroscopic system approaching thermal equilibrium, statistical mechanics – in which we deal with “ensembles” – provides a powerful framework for studying thermalization.\\
This phenomenon is related to the "ergodic" property of classical chaotic systems, which confirms statistical mechanics. Indeed, in such systems, the ensemble averages employed in statistical mechanics calculations correspond closely to the time averages observed in our studies. Even in closed quantum systems, the emergence of thermal equilibrium can be observed in nonequilibrium states; these nonequilibrium states tend to approach their thermal expectation values shortly after relaxation \cite{Srednicki:1994mfb,Deutsch1991}.\\
The concept of reaching thermal equilibrium in quantum mechanics can be formalized through the "Eigenstate Thermalization Hypothesis" (ETH) \cite{Srednicki:1994mfb,Deutsch1991}, which explains how an observable system approaches its thermal equilibrium value.According to ETH, in sufficiently complex non-integrable quantum systems, individual energy eigenstates are locally indistinguishable from thermal states at the same energy. \\
Although it is generally accepted that non-integrable models thermalize, the precise nature of thermalization may vary across different contexts. Actually, beyond the Hamiltonian that governs the system’s dynamics, thermal behavior may also depend sensitively on the initial state; within a fixed model, different initial states may lead to distinct dynamical behaviors \cite{Banuls:2010zki}.\\
According to these considerations, it is essential, before addressing the thermalization of quantum systems, to develop a thorough understanding and analysis of chaotic (non-integrable) systems in quantum mechanics. While classical chaos is well understood, the concept of quantum chaos remains ambiguous and fraught with inherent complexities.\\
Classical chaotic behavior is typified by an extreme sensitivity of phase-space trajectories to initial conditions, where two initially nearby trajectories diverge exponentially fast — a phenomenon quantitatively described by the Lyapunov exponent. In contrast, quantum chaos is less clearly defined and more challenging to comprehend, primarily owing to the absence of a well-established notion of quantum phase space.\\
Motivated by recent progress in the study of chaos in quantum many-body systems, understanding the growth and complexity of an operator under unitary evolution in the Heisenberg picture has become a prominent research focus across various subfields of physics\cite{Hosur:2015ylk,Maldacena:2015waa,Kitaev:2017awl,Roberts:2018mnp,vonKeyserlingk:2017dyr,Nahum:2017yvy}.A widely used framework to describe operator growth employs out-of-time-ordered correlators (OTOCs), which quantify the spreading of operators across both spatial and temporal domains \cite{Aleiner:2016eni,Chowdhury:2017jzb,Garttner:2016mqj,Li:2016xhw,Lin2018,Lin2018b,Gopalakrishnan:2018rfu,Khemani:2018sdn,Xu:2018xfz,Sunderhauf:2019djv,Yan:2019fbg}.\\
From a semiclassical perspective, OTOCs exhibit exponential growth attributed to the Butterfly Effect, characterized by a Lyapunov exponent that is considered to be bounded \cite{Rozenbaum:2016mmv}. This bound is saturated in certain strongly interacting models with holographic duals, such as the Sachdev-Ye-Kitaev (SYK) model \cite{Sachdev1993,Maldacena:2016hyu}. However, it is important to note that exponential growth (OTOCs) is not a universal feature of all chaotic systems \cite{Fine:2013zog,Hashimoto:2017oit,Hashimoto:2020xfr}.In chaotic quantum systems, thermalization, complexity \cite{Doroudiani:2019llj,Pirmoradian2020,RezaTanhayi:2018cyv,Khorasani:2023usq,Pirmoradian2025}, and entanglement entropy \cite{Ghasemi:2021jiy,Pirmoradian:2025nxw,Pirmoradian:2023uvt,Pirmoradian:2021wvo,Pirmoradian:2025dco} are interconnected concepts. Thermalization, the process by which a system reaches a state of equilibrium, is often linked to the growth of entanglement entropy and complexity. As a chaotic system evolves, entanglement entropy (a measure of entanglement between subsystems) increases, eventually saturating at a value related to the system's thermodynamic entropy. Complexity, often quantified using measures like circuit complexity, also tends to increase during thermalization, reflecting the system's growing internal structure and the scrambling of information.\\
Complementing existing frameworks, operator entanglement constitutes a pivotal diagnostic tool for characterizing operator complexity. Extensively examined across diverse quantum systems \cite{Prosen:2007gfp,Pizorn:2009gup,Alba:2019okd,Bertini:2019gbu,MacCormack:2020auw,Mascot:2020qep,Alba:2020npy}, operator entanglement offers profound insights into the intrinsic dynamical evolution of operators. This perspective elucidates the progressive system complexity acquisition of operators, thereby advancing our fundamental understanding of the mechanisms that govern thermalization and the scrambling of quantum information.\\
Recently, Parker and collaborators \cite{Parker:2018yvk} introduced a recursive framework for characterizing operator growth, employing a formulation rooted in the Lanczos algorithm. This approach systematically constructs an orthonormal set of states spanning the Krylov subspace, often referred to as the Krylov basis, through successive applications of the system’s Liouvillian or Hamiltonian to an initial operator or state. Within this basis, the evolution of the system is encapsulated by a sequence of time-dependent coefficients, providing a compact yet informative representation of the dynamics. Building upon this structure, the notions of Krylov complexity and Krylov entropy were defined and investigated in \cite{Barbon:2019wsy,Vasli:2023syq}, offering quantifiable measures of the spread of operators within the Krylov subspace. Due to its recursive construction and numerical stability, this methodology has seen extensive application in computational analyses \cite{Viswanath:1994}. More recently, there has been a growing interest in utilizing the Krylov-based approach as a diagnostic tool for probing signatures of quantum chaos and scrambling in many-body systems \cite{Parker:2018yvk}.\\
Recent developments have proposed that the asymptotic growth of Lanczos coefficients, which emerge in the Krylov basis representation of operator dynamics, can provide insight into the nature of quantum many-body chaos. Parker et al. \cite{Parker:2018yvk} introduced the Operator Growth Hypothesis, arguing that in spatially extended quantum systems without conservation laws and with dimensions $d>1$, the Lanczos coefficients grow linearly at large indices, $b_n\sim\alpha n$. This behavior has been linked to the spreading of operators and the rapid growth of Krylov complexity, suggesting a connection to chaotic dynamics.\\
However, subsequent studies have demonstrated that linear growth of Lanczos coefficients is not exclusive to chaotic systems. Dymarsky and Smolkin \cite{Dymarsky:2021bjq} showed that free field theories, which are integrable, can exhibit similar asymptotic growth. Bhattacharjee et al. \cite{Bhattacharjee:2022vlt} and Avdoshkin et al. \cite{Avdoshkin:2022xuw} further examined integrable quantum field theories and spin chains, finding that they may also display linear Lanczos growth. Additional works, including those by Camargo et al. \cite{Camargo:2022rnt}, have provided further examples in holographic models and interacting quantum field theories, reinforcing that while linear Lanczos growth can occur in chaotic settings, it is not a definitive signature of chaos.\\
This indicates a need to identify diagnostics that can more reliably distinguish between chaotic and integrable regimes in many-body systems. One promising candidate is the saturation value of Krylov complexity, which quantifies the dimension of the Krylov subspace effectively explored during the time evolution of an operator in a finite system. In a local quantum Ising chain, Rabinovici et al. \cite{Rabinovici:2021qqt,Rabinovici:2022beu} found that the saturation value of Krylov complexity increases as the system parameters are tuned from integrable conditions to regimes associated with chaotic dynamics, suggesting that the saturation level may serve as a practical diagnostic for distinguishing chaotic behavior in finite-size systems.\\
Recent studies have expanded the use of Krylov complexity to systems with many-body localization and near-integrable dynamics, observing that the growth and eventual saturation of complexity can separate thermalizing chaotic phases from non-ergodic phases \cite{Alishahiha:2022nhe,Baggioli:2024wbz}. These findings strengthen the argument that Krylov complexity can offer a finer tool for identifying chaos in many-body systems, especially when used in conjunction with, but not replaced by, the observation of Lanczos coefficient growth.\\
A closely related yet distinct approach to diagnosing quantum chaos involves studying energy level statistics. Quantum chaotic systems typically exhibit level repulsion and are well described by random matrix theory (RMT), leading to Wigner-Dyson distributions in the energy level spacings, while integrable systems tend to follow Poisson statistics \cite{Bohigas:1983er,DAlessio:2015qtq}. Understanding how the behavior of Krylov complexity correlates with energy level statistics can provide a more nuanced picture of chaos and integrability in quantum many-body systems. Recent works, including \cite{Chan:2018dzt,Kawabata:2021tsf}, have begun to explore these connections, highlighting that while Krylov complexity probes operator dynamics, level statistics probe spectral properties, and their interplay can yield powerful insights into the structure of chaos.\\
The remainder of this paper is organized as follows. In the next section, we provide a detailed overview of the Krylov space formalism, including the procedure for constructing the Krylov basis, calculating Lanczos coefficients, and defining Krylov complexity, with a focus on its physical interpretation in interacting many-body systems. Subsequently, we introduce the mixed-field Ising model, considering both local and non-local interaction terms, and analyze its symmetry properties as well as the parameter regimes that correspond to integrable and chaotic dynamics. Following this, we present a comprehensive numerical and analytical study of Krylov complexity alongside energy level spacing statistics, examining their interplay across the integrable-to-chaotic crossover. Finally, in the concluding section, we synthesize the findings to demonstrate how the behavior of Krylov complexity and the distribution of level spacing ratios can serve as reliable diagnostics to distinguish chaotic from integrable phases in both local and non-local variants of the Ising model.

\section{Krylov Space and Spread Complexity}
Consider a time-independent quantum system characterized by a Hamiltonian $H$. The unitary time evolution of a pure state $\left| \psi(t) \right\rangle$ is governed by the Schrödinger equation:
\begin{equation}\label{eq.1}
i\left.\partial_t\left|\psi\left(t\right)\right.\right\rangle=H\left.\left|\psi\left(t\right)\right.\right\rangle
\end{equation}
The formal solution is $\left.\left|\psi\left(t\right)\right.\right\rangle=e^{-iHt}\left.\left|\psi\left(0\right)\right.\right\rangle$.

\noindent This result can conventionally be expressed as a power series expansion in terms of states obtained by successive operations of the Hamiltonian on the initial state at $t=0$. 

\begin{figure}[h]
    \centering
    \begin{minipage}{0.48\textwidth}
        \centering
        \includegraphics[width=\textwidth]{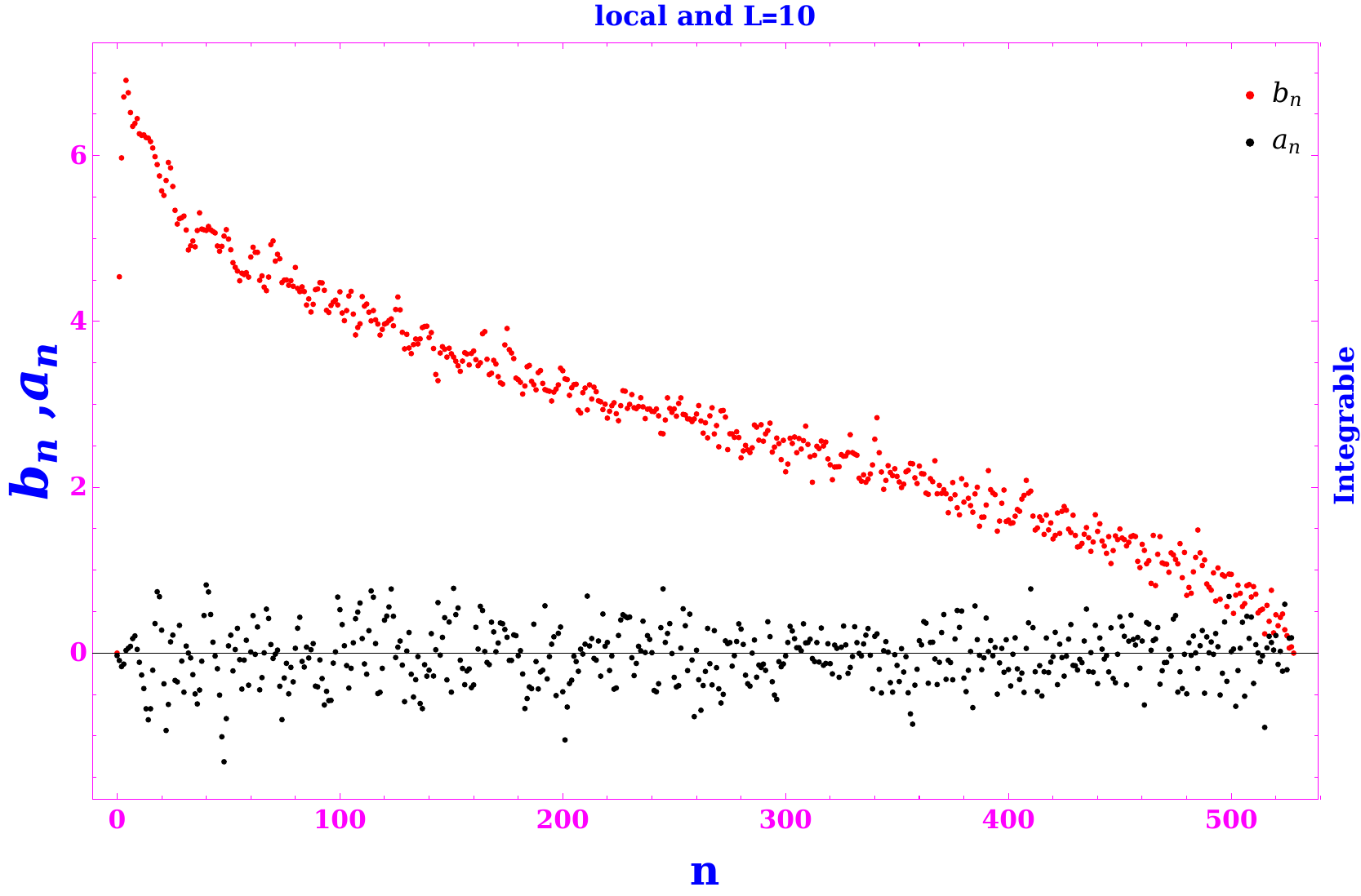}
    \end{minipage}
    \hfill
    \begin{minipage}{0.48\textwidth}
        \centering
        \includegraphics[width=\textwidth]{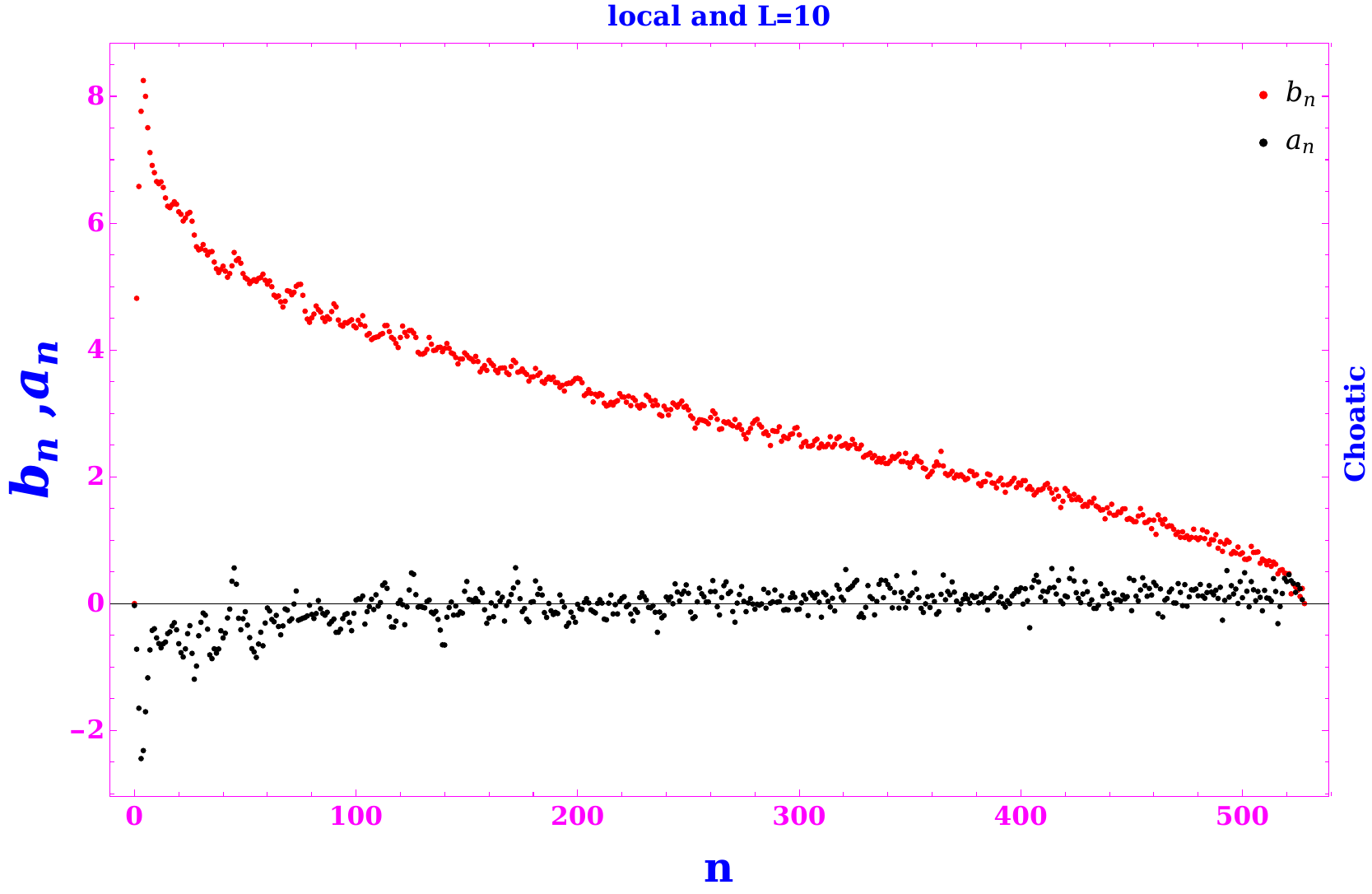}
    \end{minipage}
    \vspace{0.5cm}
    \begin{minipage}{0.48\textwidth}
        \centering
        \includegraphics[width=\textwidth]{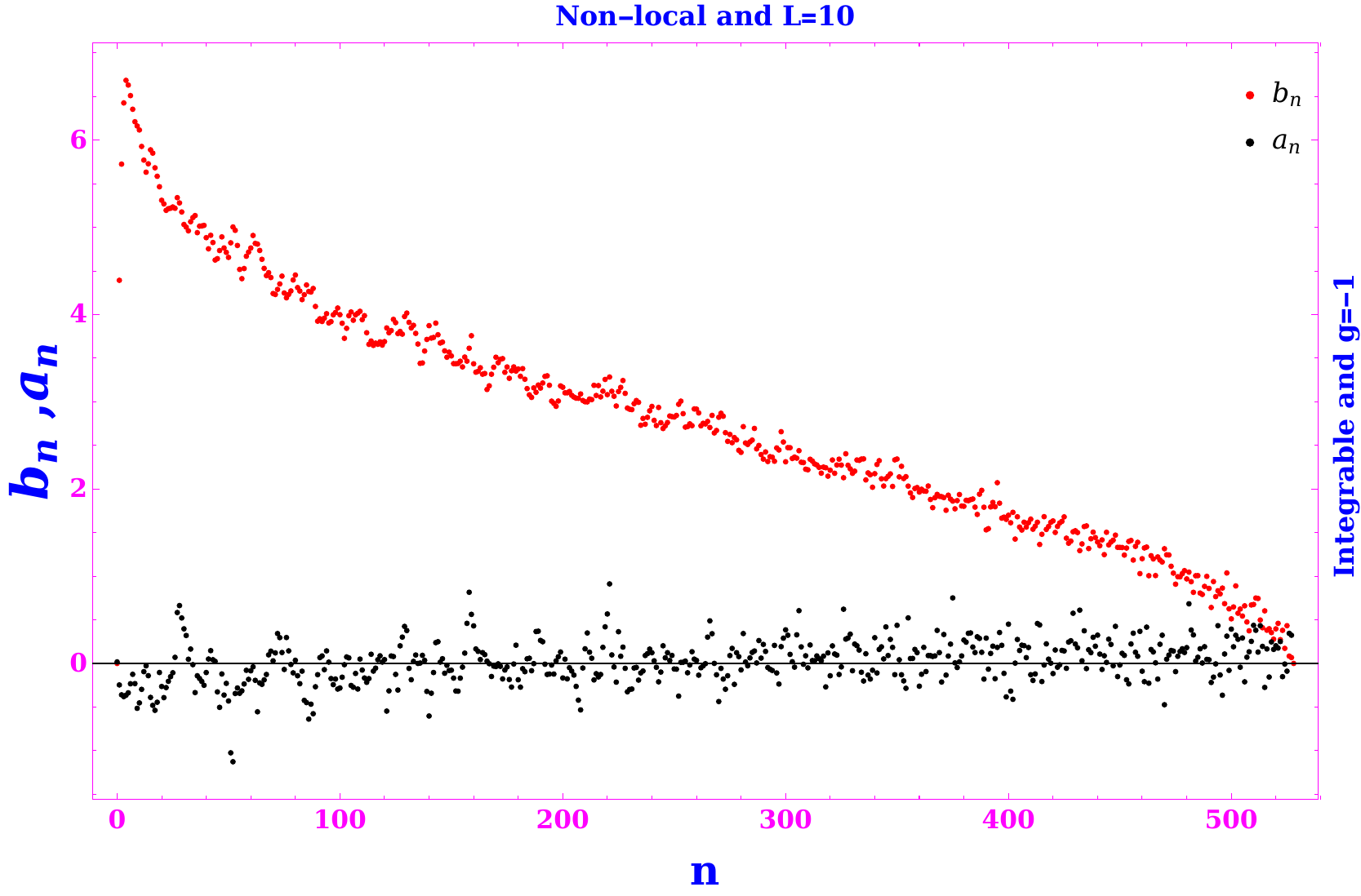}
    \end{minipage}
    \vspace{0.5cm}
    \begin{minipage}{0.48\textwidth}
        \centering
        \includegraphics[width=\textwidth]{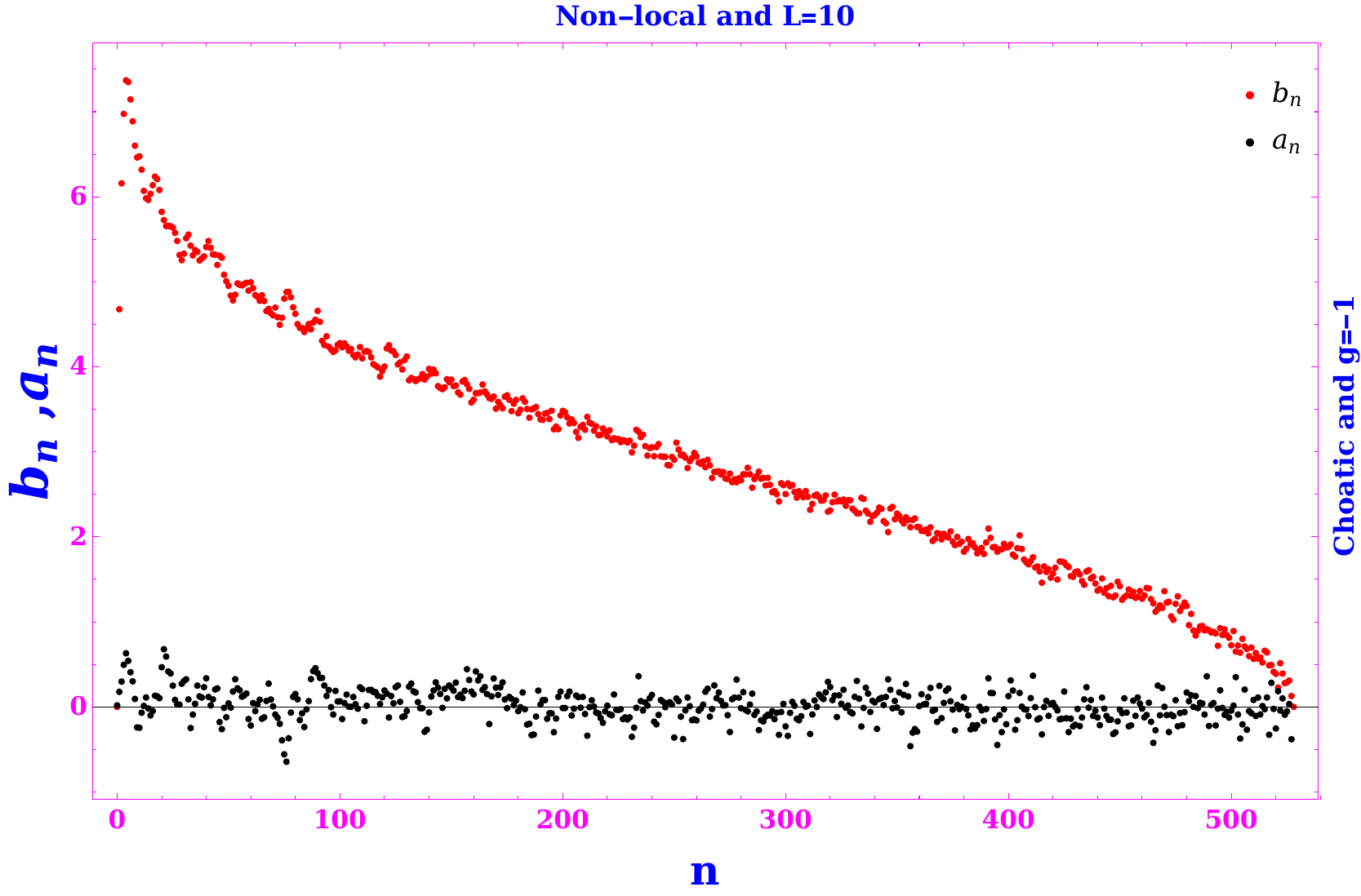}
    \end{minipage}
    \caption{The Lanczos coefficients for the local version of the mixed-field Ising model (top plots) and the non-local version (bottom plot), in both integrable and chaotic forms are shown.}
    \label{Fig.1}
\end{figure}

\noindent That is:
\begin{equation}\label{eq.2}
\left.\left|\psi\left(t\right)\right.\right\rangle=\sum_{n=0}^{\infty}\frac{\left(-it\right)^n}{n!}\left.\left|\psi_n\right.\right\rangle
\end{equation}\\
where the states:
\begin{equation}\label{eq.3}
\left.\left|\psi_n\right.\right\rangle:=H^n\left.\left|\psi\left(0\right)\right.\right\rangle. 
\end{equation}\\
These states span what is known as the Krylov space, which is a subspace of the full Hilbert space. Although these states generally are neither orthogonal nor properly normalized, to construct a set of orthonormal basis vectors $\left.\left|K_n\right.\right\rangle$, known as Krylov's basis, we employ the Lanczos algorithm \cite{Muck:2022xfc}, which is based on the Gram-Schmidt orthogonalization process. The Lanczos recursion is given by:
\begin{equation}\label{eq.4}
    \left.\left|A_{n+1}\right.\right\rangle=\left(H-a_n\right)\left.\ \left|K_n\right.\right\rangle-b_n\left.\ \left|K_{n-1}\right.\right\rangle
\end{equation}
Here, $a_n$ and $b_n$ are the Lanczos coefficients, defined as:
\begin{equation}\label{eq.5}
    a_n=\left\langle K_n\middle| H\middle| K_n\right\rangle,\ \ \ b_n = \langle A_n \mid A_n \rangle^{1/2}
\end{equation}
The initial and normalized conditions of Krylov's basic elements can be presented as follows:
\begin{equation}\label{eq.6}
\left.\left|K_0\right.\right\rangle=\left.\left|\psi\left(0\right)\right.\right\rangle\ \ ,\ \ \ \left.\left|K_n\right.\right\rangle=b_n^{-1}\left.\ \left|A_n\right.\right\rangle\ \ \ ,b_0 := 0
\end{equation}
Using this framework, the Lanczos algorithm can be rewritten as:
\begin{equation}\label{eq.7}
 H\left.\ \left|K_n\right.\right\rangle=a_n\left.\ \left|K_n\right.\right\rangle+b_{n+1}\left.\ \left|K_{n+1}\right.\right\rangle+b_n\left.\ \left|K_{n-1}\right.\right\rangle
\end{equation}
This expression shows that the Hamiltonian takes a tri-diagonal form in Krylov's basis. In an infinite-dimensional Hilbert space, the Krylov space is also infinite-dimensional. In matrix form:
\begin{equation}\label{eq.8}
\left\langle K_m\middle| H\middle| K_n\right\rangle=a_n\delta_{m,n}+b_{n+1}\delta_{m,n+1}+b_n\delta_{m,n-1}
\end{equation}
For a finite-dimensional Hilbert space, the Krylov subspace will be finite-dimensional, with its dimension determined by the initial state and the Hamiltonian's structure. The Lanczos process terminates when $b_n=0$ for some $n$, indicating the generated vectors have become linearly dependent.\\
In Fig \ref{Fig.1}, the Lanczos coefficients for the mixed-field Ising model are computed and presented. The behavior of the Lanczos coefficients $b_n$ provides insight into the operator growth
dynamics across different regimes. As shown in  Fig. \ref{Fig.1}, the Lanczos coefficients $b_n$ for the chaotic parameter set exhibit an approximately smooth, nearly linear growth with smaller point-to-point fluctuations, whereas the integrable case shows more pronounced oscillations and larger local variance.
\\
The time-evolved state can be decomposed in the Krylov basis as:
\begin{equation}\label{eq.9}
\left.\left|\psi\left(t\right)\right.\right\rangle=\sum_{\ n}{\psi_n\left(t\right)}\left.\left|K_n\right.\right\rangle
\end{equation}
Where ${\ \psi}_n\left(t\right):=\left\langle K_n\middle|\psi\left(t\right)\right\rangle$ are functions as follows:
\begin{equation}\label{eq.10}
\sum_{n}{\left|\psi_n\left(t\right)\right|^2=}\sum_{n}{p_n\left(t\right)=1}   
\end{equation}
By inserting Eq.\ref{eq.7} into Eq.\ref{eq.1}, the Schrödinger equation reduces to a recursion relation for the Krylov coefficients:
\begin{equation}\label{eq.11}
i\partial_t\psi_n\left(t\right)=a_n\psi_n\left(t\right)+b_{n+1}\psi_{n+1}\left(t\right)+b_n\psi_{n-1}\left(t\right) 
\end{equation}\\
For a given initial state $\left.\left|\psi\left(0\right)\right.\right\rangle$ with the initial condition $\psi_n\left(0\right)=\delta_{n0}$, one defines the Krylov complexity as:
\begin{equation}\label{eq.12}
C\left(t\right):=\sum_{n}{{n\ p}_n\left(t\right)=}\sum_{n}{n\left|\psi_n\left(t\right)\right|}^2
\end{equation}\\
Which serves as a measure of the effective spread of the initial state along the Krylov chain under unitary evolution. This quantity encodes the growth of operator or state complexity in a one-dimensional effective representation of the dynamics and is of particular interest when analyzing chaotic versus integrable behavior in many-body quantum systems.\\

\section{The mixed-field Ising Model}
The mixed-field Ising model extends the standard Ising model in statistical mechanics by including both transverse and longitudinal magnetic fields acting on spin sites, allowing the system to interpolate between integrable and quantum chaotic dynamics depending on the field strengths \cite{Craps:2019rbj,Caceres:2024mec}. The model retains local nearest-neighbor interactions, and its Hamiltonian takes the form:
\begin{equation}\label{eq.13}
H=-\sum_{i=1}^{L-1}{s_i^zs_{i+1}^z-\sum_{i=1}^{L}\left({h_xs}_i^x{-h_zs}_i^z\right)}
\end{equation}\\
Where $s_i^x$ and $s_i^z$ denote spin-$\frac{1}{2}$ operators on site $i$. These are constructed explicitly as:
\begin{equation}\label{eq.14}
s_i^k=(\mathbb{I}_2)^{\otimes (i-1)} \otimes \sigma_k \otimes (\mathbb{I}_2)^{\otimes (L-1)}, \quad k\in\left\{x,y,z\right\}
\end{equation}\\
Here, $\sigma_k$ are Pauli matrices, and the Hilbert space dimension is $d=2^L$. \\
To explore the effects of long-range interactions on quantum chaos, we consider a non-local
extension of the mixed-field Ising model, often referred to as the fast-entangling spin model
\cite{Belyansky:2020bia}. The Hamiltonian includes an all-to-all interaction term in the $x$-component of
the spins,
\begin{equation}
H_{\text{non-local}} = -\frac{g}{\sqrt{L}} \sum_{i<j} s_i^x s_j^x ,
\end{equation}
where $g$ is a dimensionless coupling strength independent of the system size $L$.
The normalization factor $1/\sqrt{L}$ ensures extensivity in the thermodynamic limit.
Since the number of interaction terms scales quadratically with system size ($\sim L^2$),
this normalization guarantees a linear scaling of the energy per site with $L$.

This model is known to induce rapid entanglement growth and is frequently employed in
studies of scrambling and quantum chaos in systems with collective interactions. In our
analysis, we observe that even modest non-local couplings (specifically $|g| > 0.25$) are
sufficient to produce clear signatures of chaos, as demonstrated by the spectral statistics
shown in Fig.~5. This behavior stands in contrast to the local Ising model, where the onset
of chaos typically requires careful tuning of transverse and longitudinal fields,
highlighting the role of non-locality in amplifying chaotic dynamics.
For specific field configurations, the model transitions between chaotic and integrable regimes. For example, numerical studies indicate that $\left(h_x,h_z\right)= (-1.05, 0.5)$ yields signatures of quantum chaos, while $\left(h_x,h_z\right)=(-1,0)$ corresponds to an integrable point \cite{Craps:2019rbj,Caceres:2024mec}:
\begin{equation}\label{eq.15}
\left(h_x,h_z\right) = 
\begin{cases}
    \left(-1.05,0.5\right)\ \ Chaotic\ State \\
   \left(-1,0\right)\ \ \ \ \ \ \ \ Integrable\ State
\end{cases}
\end{equation}\\
The model possesses parity symmetry, with the Hamiltonian commuting with the parity Operator $\hat{\Pi}$. This operator can be expressed as:
\begin{equation}\label{eq.16}
\hat{\Pi}= 
\begin{cases}
   {\hat{p}}_{1,L}~{\hat{p}}_{2,L-1\ }\ldots{\hat{p}}_{\frac{L}{2},\frac{L+2}{2}}\ \ \ \ \ \ \ when\ L\ is\ even \\
    {\hat{p}}_{1,L}~{\hat{p}}_{2,L-1\ }\ldots{\hat{p}}_{\frac{L-1}{2},\frac{L+3}{2}}\ \ \ \ \ when\ L\ is\ odd
\end{cases}
\end{equation}\\
Where the site permutation operator is defined by:
\begin{equation}\label{eq.17}
{\hat{p}}_{i,j}=\frac{1}{2}\left(\mathbb{I}_d+s_i^xs_j^x+s_i^ys_j^y+s_i^zs_j^z\right)
\end{equation}\\
And it exchanges the spin states at sites $i$ and $j$. Conceptually, $\hat{\Pi}$ acts by reflecting the spin configuration about the chain's center, e.g.,
\begin{equation*}
\hat{\Pi}\left|\left.\uparrow\uparrow\uparrow\downarrow\right\rangle\right.=\left|\left.\downarrow\uparrow\uparrow\uparrow\right\rangle\right.,
\end{equation*}\\
And enables the construction of eigenstates with definite parity:
\begin{equation*}
\left.\left|\psi_\pm\right.\right\rangle=\frac{1}{\sqrt2}\left(\left|\left.\uparrow\uparrow\uparrow\downarrow\right\rangle\right.\pm\left|\left.\downarrow\uparrow\uparrow\uparrow\right\rangle\right.\right),
\end{equation*}\\
With $\hat{\Pi}\left|\left.\psi_\pm\right\rangle\right.=\pm\left|\left.\psi_\pm\right\rangle,\right.$ having eigenvalues $\pm1$ \cite{Rabinovici:2022beu}.\\
To explore the effects of long-range interactions in a disordered transverse field environment, we also consider the fast entangling spin model introduced in \cite{Belyansky:2020bia}, whose Hamiltonian is given by:
\begin{equation}\label{eq.18}
H=H_{local}-\frac{g}{\sqrt L}\sum_{i<j}{s_i^zs_j^z},
\end{equation}\\
Where $H_{local}$ denotes the local mixed-field Ising Hamiltonian. This extension enables the study of entanglement spreading and level statistics in systems that incorporate both local and collective interaction structures, offering a broader landscape for investigating the signatures of chaos and integrability in many-body quantum systems.\\

\section{Krylov Complexity and energy level spacing}
In this section, we analyze the role of Krylov complexity in diagnosing quantum chaos in many-body systems over long timescales. Krylov complexity, defined as the expectation value of the Krylov number operator, provides a quantitative measure of the operator’s spreading in Krylov space \cite{Parker:2018yvk}:
\begin{equation}\label{eq.19}
C(t)=\left \langle \Psi(t)| N | \Psi(t) \right \rangle
\end{equation}\\
where the Krylov number operator in the Krylov basis takes the form:
\begin{equation}\label{eq.20}
N=\sum_{n=0}^{D_{\Psi}-1}n\left| n \right\rangle \left \langle n \right |
\end{equation}\\
To track the time dependence of complexity, one studies the temporal evolution of this expectation value \cite{Rabinovici:2022beu}. We will show that the long-time behavior of $C(t)$ can reveal signatures of chaos or integrability in the underlying dynamics.\\
Given an initial state:
\begin{equation*}
\left|\left.\psi_0\right\rangle\right.=\sum_{j=1}^{D_\psi}{c_j\left|\left.E_j\right\rangle\right.}
\end{equation*}\\
the Krylov complexity can be expressed as:
\begin{equation}\label{eq.21}
C(t)=Tr(\rho_{DE}N)+\sum_{j=1}^{D_{\psi}}e^{i\omega_{jk}t}c_{j}^{*}c_{k}N_{jk},
\end{equation}\\
Where, $\omega_{jk}=E_{j}-E_{k},N_{jk}=\langle E_{j}| N |E_{k} \rangle$ and the diagonal ensemble density matrix is defined by:
\begin{equation}\label{eq.22}
\rho_{DE}=\sum_{j=1}^{D_\psi}{\left|c_j\right|^2\left|\left.E_j\right\rangle\left\langle\left.E_j\right|\right.\right.}.
\end{equation}\\ 
Throughout, we restrict to the symmetry sector relevant to the initial state, ensuring that the Krylov space and $\rho_{DE}$ are confined to the block associated with conserved charges or parity.\\
In our analysis of Krylov complexity, we employ an initial state prepared at infinite
temperature ($\beta = 0$), corresponding to the maximally mixed state within the relevant
symmetry sector. This choice ensures uniform sampling of the Hilbert space and provides
a generic probe of the system's dynamics, independent of specific initial configurations.
For chaotic systems, the growth profile and saturation of Krylov complexity are expected
to be universal across a broad class of initial states, whereas in integrable regimes, the
behavior may exhibit stronger dependence on the initial condition. While simpler product
states, such as polarized spin configurations, show qualitatively similar trends—particularly
the characteristic peak preceding saturation in chaotic phases—the thermal initial state
offers a cleaner diagnostic by mitigating biases arising from special initial configurations.
We note that although quantitative measures of complexity may vary with the choice of
initial state, the emergence of a distinct peak in $C(t)$ and its saturation level remain
robust indicators of chaotic dynamics across a wide range of initial conditions.

\begin{figure}
\centering
    \begin{minipage}{0.48\textwidth}
        \centering
        \includegraphics[width=\textwidth]{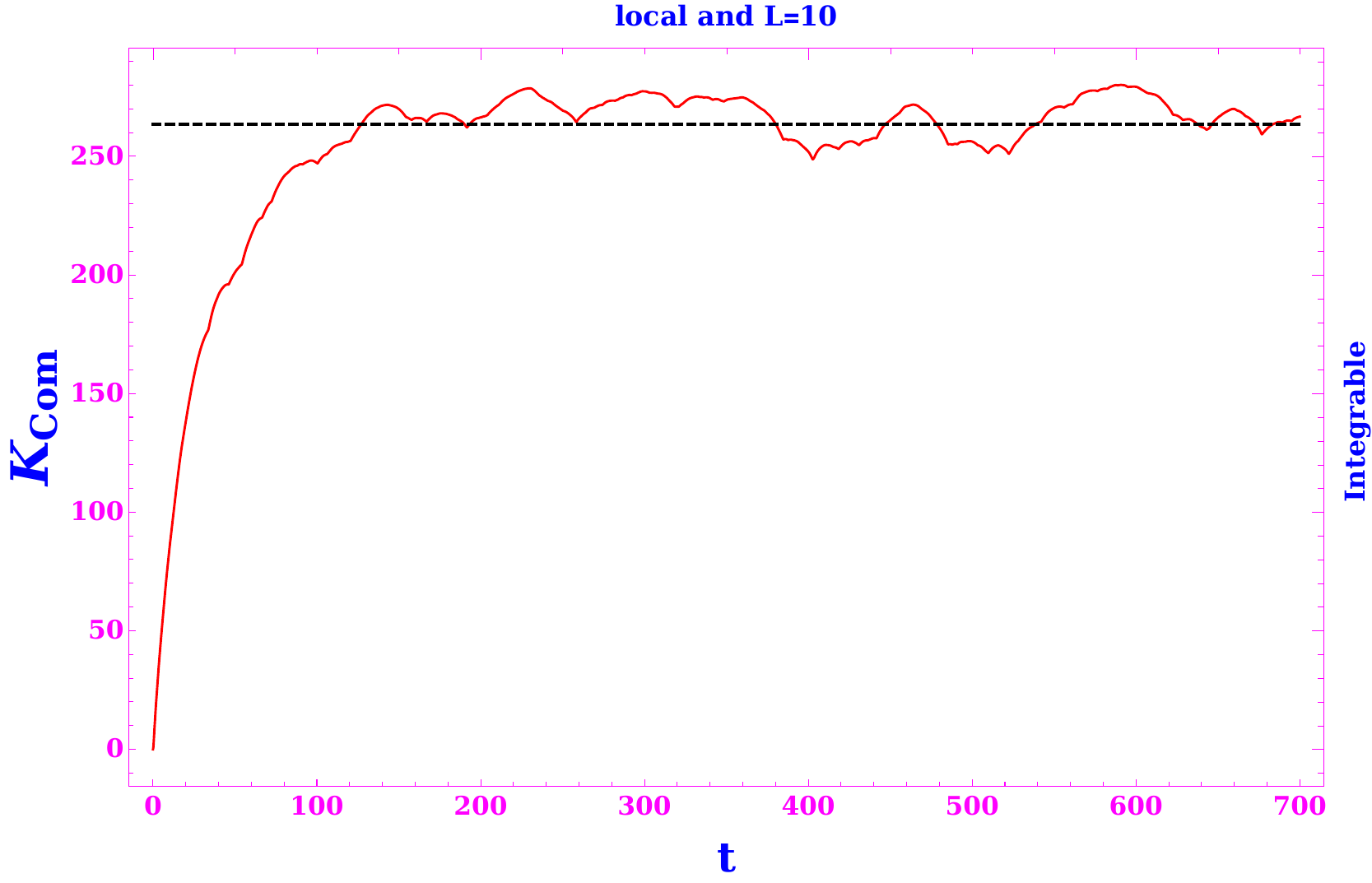}
    \end{minipage}
    \hfill
    \begin{minipage}{0.48\textwidth}
        \centering
        \includegraphics[width=\textwidth]{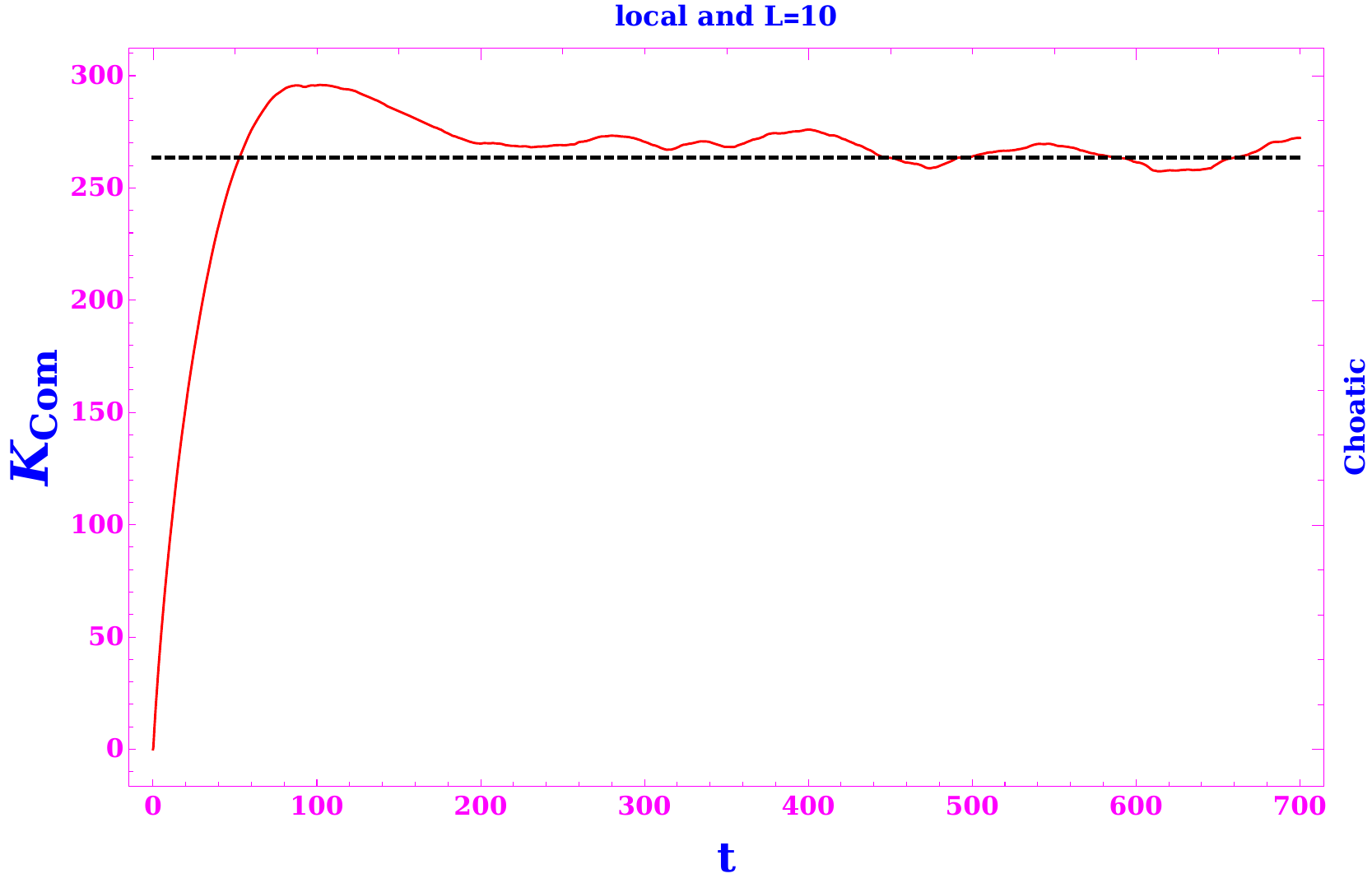}
    \end{minipage}
    \vspace{0.5cm}
    \begin{minipage}{0.48\textwidth}
        \centering
        \includegraphics[width=\textwidth]{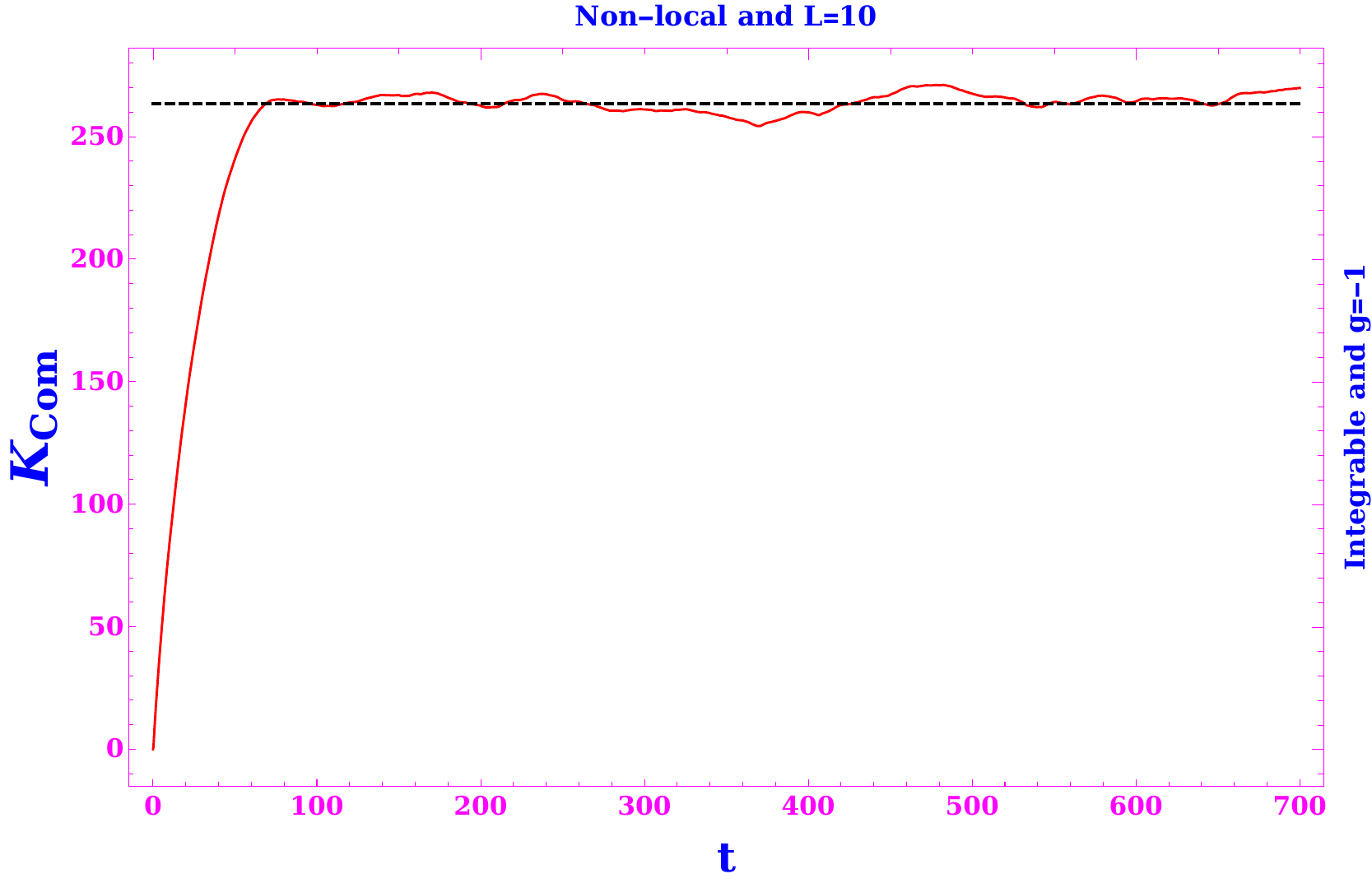}
    \end{minipage}
    \vspace{0.5cm}
    \begin{minipage}{0.48\textwidth}
        \centering
        \includegraphics[width=\textwidth]{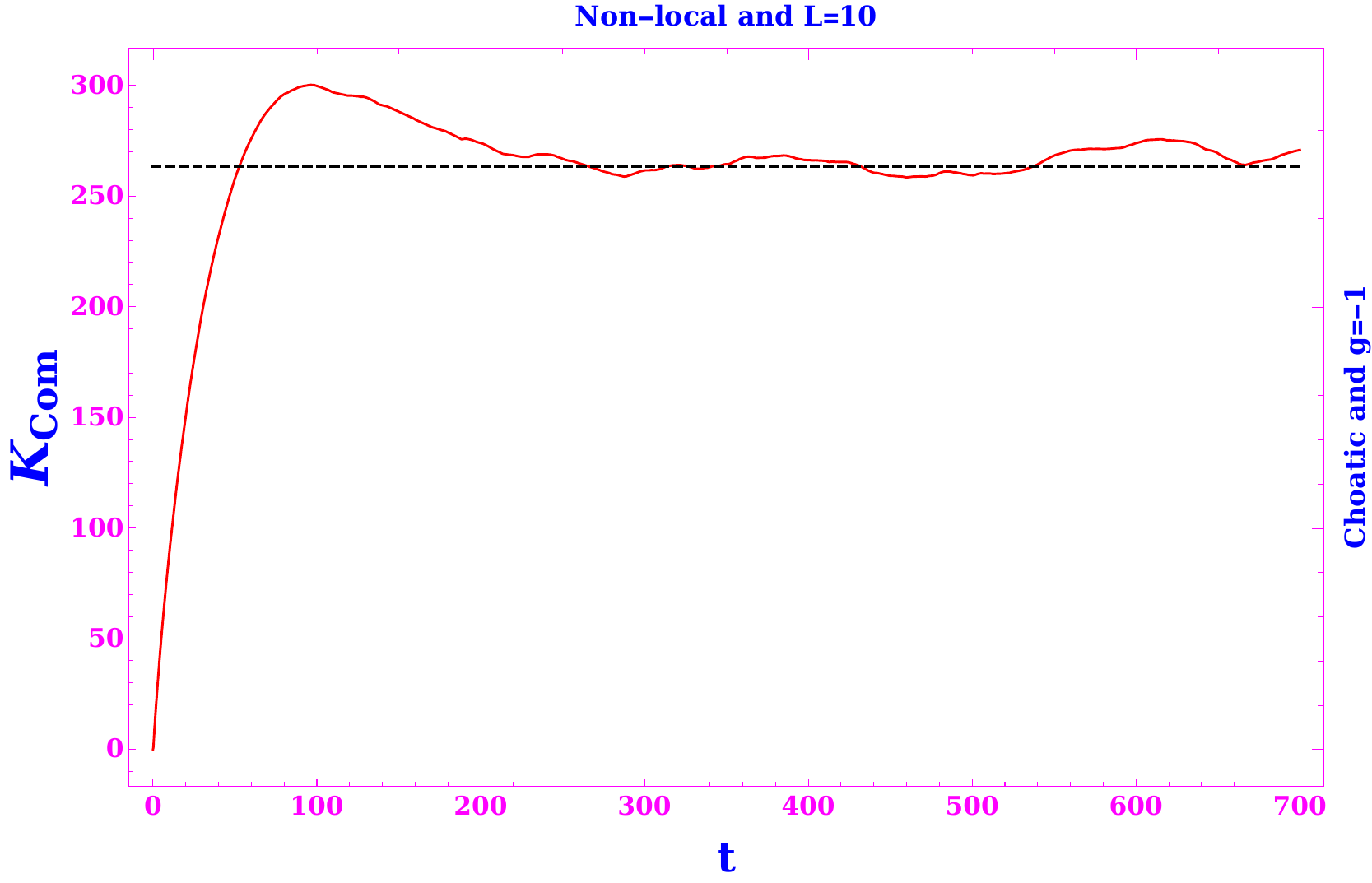}
    \end{minipage}
    \caption{We examine the time dependence of the Krylov complexity for the initial state defined in Eq.\ref{eq.26}, evaluated at $\beta=0$, which corresponds to the infinite-temperature limit. In the regime where the system exhibits chaotic dynamics, we observe that the Krylov complexity increases rapidly, reaches a pronounced maximum, and subsequently saturates due to the finite dimension of the accessible Hilbert space. The black dashed line denotes the infinite-time average of the Krylov complexity.}
    \label{Fig.2}
\end{figure}

\begin{figure}
\centering
    \begin{minipage}{0.48\textwidth}
        \centering
        \includegraphics[width=\textwidth]{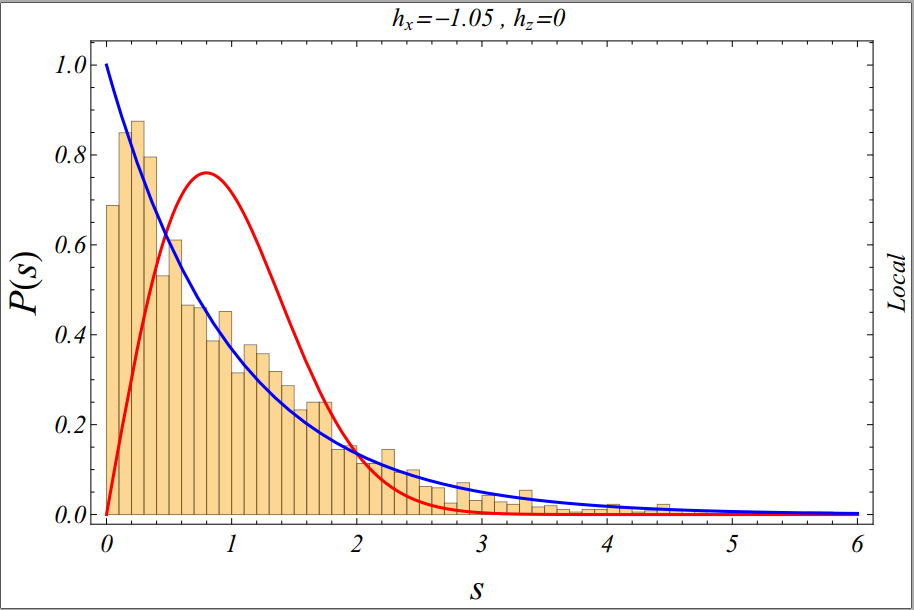}
    \end{minipage}
    \hfill
    \begin{minipage}{0.48\textwidth}
        \centering
        \includegraphics[width=\textwidth]{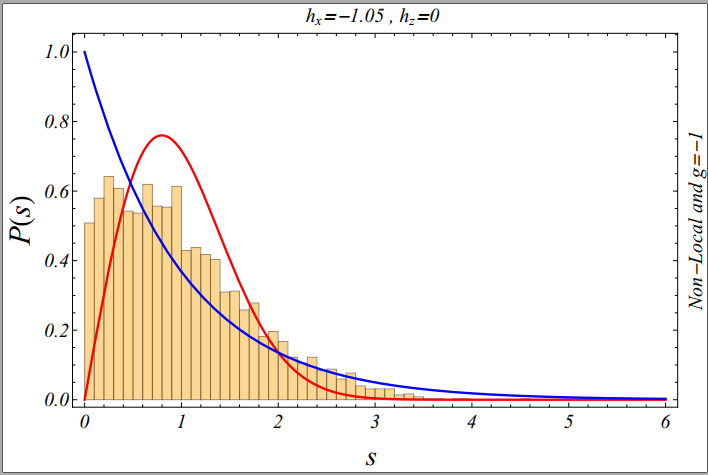}
    \end{minipage}
    \vspace{0.5cm}
    \begin{minipage}{0.48\textwidth}
       \centering
        \includegraphics[width=\textwidth]{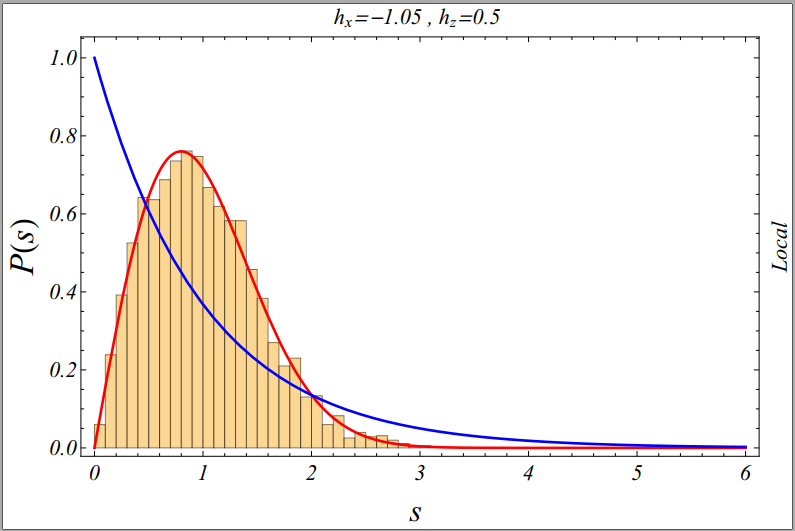}
    \end{minipage}
    \hfill
    \begin{minipage}{0.48\textwidth}
        \centering
        \includegraphics[width=\textwidth]{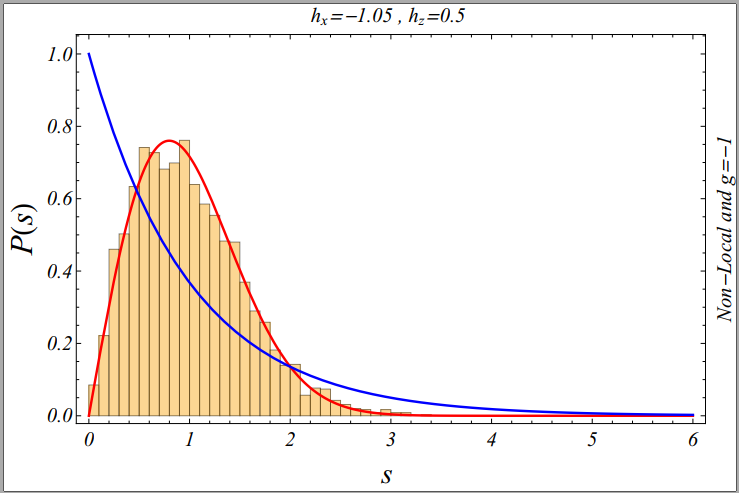}
    \end{minipage}
    \caption{Energy level spacing distributions are presented for the positive parity sector of both the local and non-local Ising models at $h_z= 0$ and $0.5$, Also, in these plots, $L=13$ is considered. The blue curve corresponds to the Poisson distribution, illustrating the behavior expected in integrable regimes, while the red curve represents the Wigner-Dyson distribution, characteristic of chaotic dynamics.}
    \label{Fig.3}
\end{figure}

\begin{figure}
\centering
    \begin{minipage}{0.48\textwidth}
        \centering
        \includegraphics[width=\textwidth]{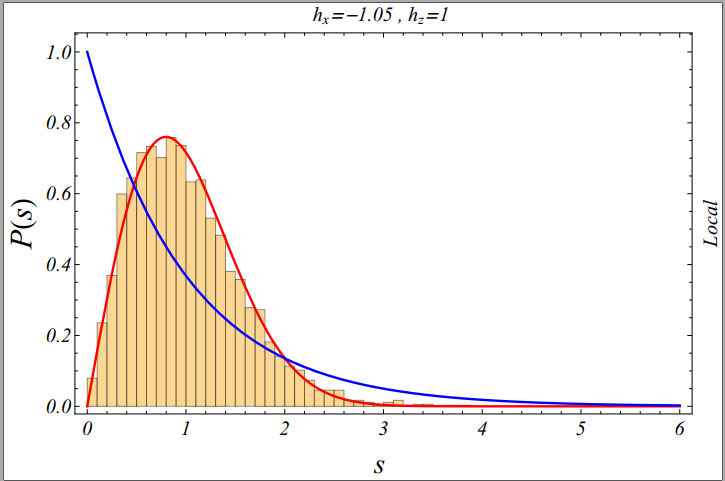}
    \end{minipage}
    \hfill
    \begin{minipage}{0.48\textwidth}
        \centering
        \includegraphics[width=\textwidth]{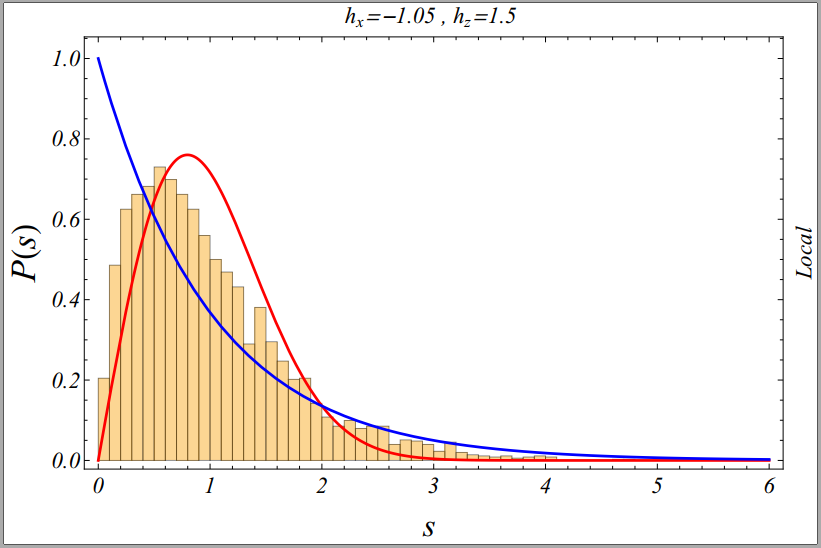}
    \end{minipage}
    \vspace{0.5cm}
    \begin{minipage}{0.48\textwidth}
       \centering
        \includegraphics[width=\textwidth]{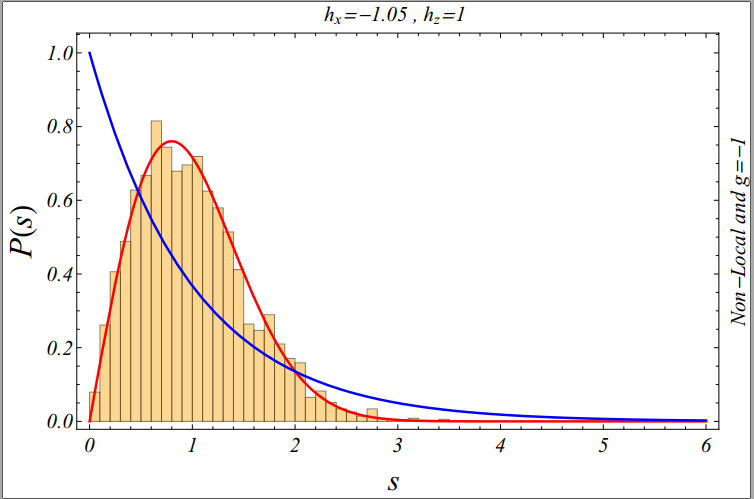}
    \end{minipage}
    \hfill
    \begin{minipage}{0.48\textwidth}
        \centering
        \includegraphics[width=\textwidth]{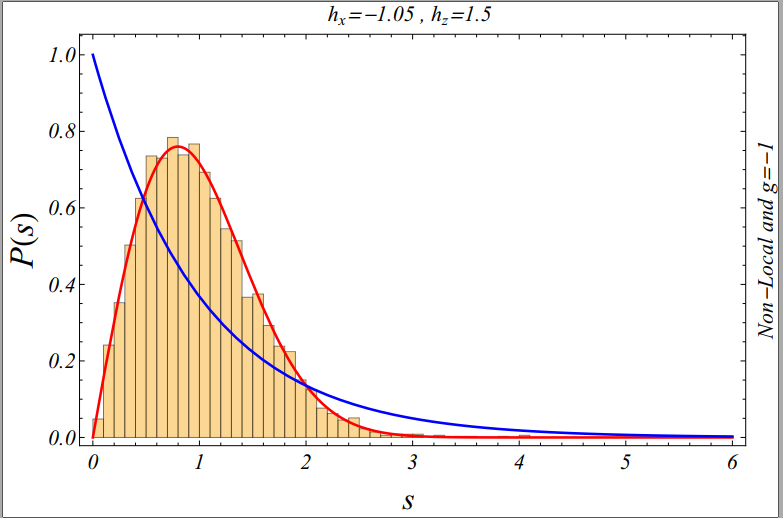}
    \end{minipage}
    \caption{Energy level spacing distributions in the positive parity sector of the local and non-local Ising models at $h_{z}=1,1.5$, Also, in these plots, $L=13$ is considered. For the local Ising model, increasing $h_{z}$ suppresses chaotic signatures, shifting the level spacing statistics from Wigner-Dyson (red) toward Poisson (blue) distributions. In contrast, the non-local Ising model exhibits an enhanced tendency toward chaos under increasing $h_{z}$, reflected in a transition from Poisson-like to Wigner-Dyson statistics across specific parameter regimes.}
    \label{Fig.4}
\end{figure}

\begin{figure}[h]
    \centering
    \begin{minipage}{0.48\textwidth}
        \centering
        \includegraphics[width=\textwidth]{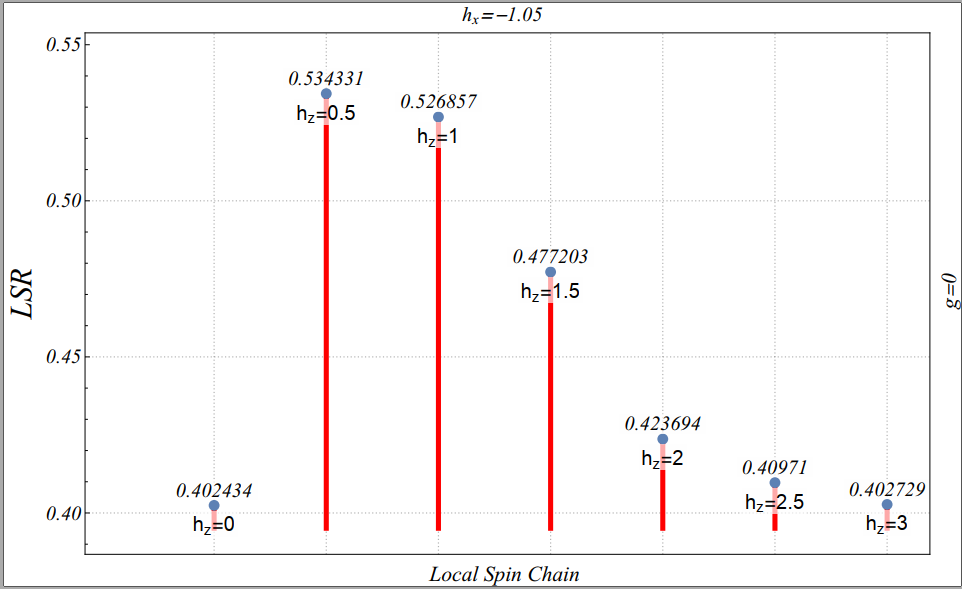}
    \end{minipage}
    \hfill
    \begin{minipage}{0.48\textwidth}
        \centering
        \includegraphics[width=\textwidth]{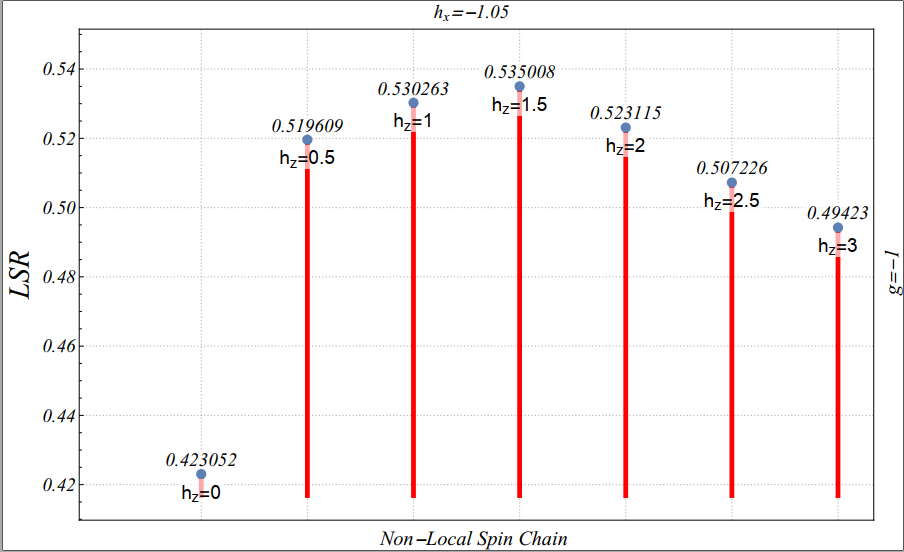}
    \end{minipage}
    \vspace{0.65cm}
    \begin{minipage}{0.58\textwidth}
        \centering
        \includegraphics[width=\textwidth]{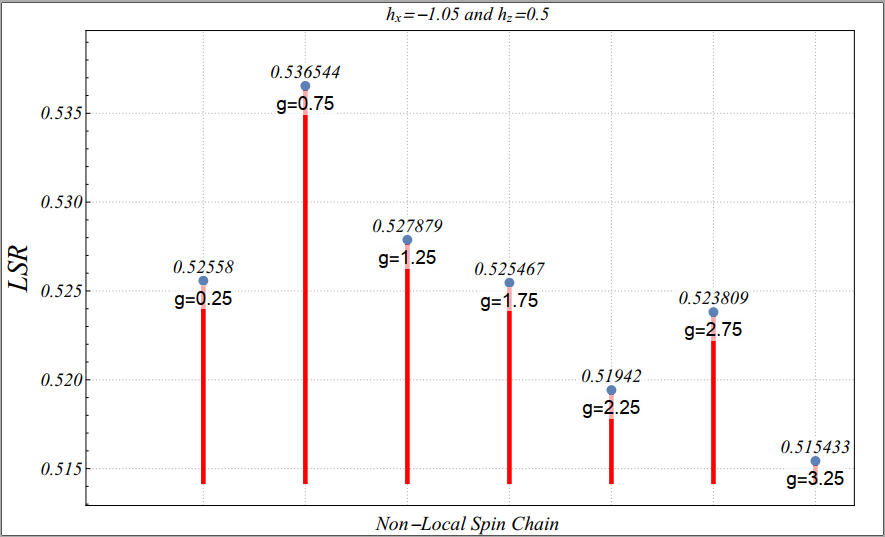}
    \end{minipage}
    \caption{The panels display the ratio of adjacent energy level spacings in the mixed-field Ising model with non-local interactions. In these plots, we have considered $L=13$. The top-left panel shows the distribution of level spacing ratios for the transverse-field Ising model $(g = 0)$ under various values of the longitudinal field $h_{z}$. The top-right panel presents the results for the mixed-field Ising model with fixed $g = -1$ and varying $h_{z}$. The bottom panel shows the level spacing ratio distributions for fixed $h_{z}=0.5$ while varying the non-local interaction strength $g$.}
    \label{Fig.5}
\end{figure}

The infinite-time average of Krylov complexity is then:
\begin{equation}\label{eq.23}
\bar{C}=\lim_{T \to \infty}\frac{1}{T}\int_{0}^{T}\langle N(t) \rangle dt=Tr(\rho_{DE}N)
\end{equation}\\
Analyzing the saturation value and the approach to saturation of Krylov complexity in a finite many-body systems offers insight into integrable versus chaotic dynamics. In chaotic systems, a pronounced peak in $C(t)$ often precedes saturation, indicating rapid operator growth before equilibration \cite{Parker:2018yvk,Rabinovici:2022beu}.\\
For thermal states, we can consider the thermal density matrix
\begin{equation}\label{eq.24}
\rho_{th}=\frac{e^{-\beta h}}{Z\left(\beta\right)}\ ,Z\left(\beta\right)=Tr\left(e^{-\beta h}\right)  
\end{equation}\\
Here $\beta$ is the inverse temperature, which is determined in the equation $Tr(\rho_{th}N)=\langle \psi_{0} | H | \psi_{0} \rangle=E_{0}$ and thermal Krylov complexity is defined by:
\begin{equation}\label{eq.25}
C_{th}=Tr(\rho_{th}N)
\end{equation}\\
which may approximate the saturation value of Krylov complexity if $N$ behaves as an ETH operator. However, in chaotic systems, it is generally observed that while local observables equilibrate quickly, the growth of Krylov complexity can continue beyond the thermalization timescale, with the late-time saturation of $C(t)$ typically exceeding $C_{th}$.\\
To study these effects explicitly, we can consider the thermal initial state:
\begin{equation}\label{eq.26}
\left.\left|\psi_0\right.\right\rangle=\frac{1}{\sqrt{Z\left(\beta\right)}}\sum_{j=1}^{D_\psi}{e^{-\frac{1}{2}\beta E_j}\left.\left|E_j\right.\right\rangle}  
\end{equation}\\
The mixed-field Ising model used here possesses parity symmetry, allowing the energy eigenstates to be classified into even and odd parity sectors, which simplifies the numerical calculation of Lanczos coefficients and Krylov complexity. For numerical studies, we compute $C(t)$ for the thermal initial state at $\beta=0$ as a representative case. As demonstrated in Fig.\ref{Fig.2}, chaotic systems exhibit a distinct peak in Krylov complexity before saturation, illustrating rapid operator spreading.\\
In isolated quantum systems, the signatures of chaos are encoded in the spectrum of the Hamiltonian. Specifically, the energy level spacings in chaotic systems typically follow a Wigner-Dyson distribution, while integrable systems exhibit Poissonian statistics. Explicitly, In an isolated quantum system defined by its Hamiltonian, along with specific boundary conditions or initial state preparations, it is natural to expect that key indicators of its chaotic nature are encoded in the eigen structure of its states.\\
In fact, the eigenvalues of chaotic Hamiltonians exhibit distinct statistical properties compared to integrable models. More precisely, the energy level spacing of the eigenvalues of a chaotic quantum system follows a Wigner-Dyson distribution, while integrable systems follow a Poisson distribution.
\begin{equation}\label{eq.27}
P_{Poisson}\left(s\right)=e^{-s}\ \ \ ,P_{WD}\left(s\right)=\frac{\pi}{2}{se}^{-\frac{\pi}{4}s^2}
\end{equation}\\ 
where $s$ denotes the normalized level spacing. These level spacing statistics serve as robust indicators for distinguishing chaotic behavior from integrability, applicable across local and non-local Ising models under transverse and longitudinal fields. As shown in Figs.\ref{Fig.3} and \ref{Fig.4}, chaotic regimes manifest clear adherence to Wigner-Dyson statistics, while integrable phases are characterized by Poissonian distributions.\\
\\
In Fig.\ref{Fig.4}, we analyze the distribution of energy level spacings within the positive parity sector for both the local and non-local variants of the Ising model under longitudinal field values $h_{z}=1$ and $1.5$. For the local Ising chain, we find that increasing $h_{z}$ progressively suppresses signatures of chaos, reflected in the gradual shift of the level spacing statistics from a Wigner-Dyson distribution, indicative of level repulsion in chaotic systems, toward a Poisson distribution, which signals the emergence of integrable behavior. In contrast, in the non-local version of the Ising model, we observe that increasing the longitudinal field can, within specific parameter windows, enhance indicators of chaos, causing the spectral statistics to evolve from a Poisson-like distribution toward Wigner-Dyson behavior.\\
However, as the longitudinal field becomes sufficiently large, the non-local model's spectral characteristics begin to resemble those of the local model, consistent with the suppression of chaotic features (see the right panel of Fig.\ref{Fig.5}). In the figures, the Poisson distribution is depicted in blue and the Wigner-Dyson distribution in red, illustrating the crossover in spectral statistics as the field strength varies.\\
\subsection{Correlation between Spectral Statistics and Krylov Complexity}

The diagnostic power of our analysis is enhanced by comparing spectral statistics with the
dynamical behavior of Krylov complexity. These two probes offer complementary
perspectives: while level spacing statistics reflect the Hamiltonian’s spectral correlations,
Krylov complexity tracks the real-time spreading of operators in Hilbert space. In chaotic
regimes, where the average level spacing ratio $\bar{r}$ approaches the Gaussian Orthogonal
Ensemble (GOE) value of $\bar{r} \approx 0.53$ and the spacing distribution follows the
Wigner--Dyson form, the Krylov complexity $C(t)$ exhibits characteristic rapid growth, a
distinct peak, and subsequent saturation at a relatively high plateau. Conversely, in
integrable phases where $\bar{r} \approx 0.39$ and Poisson statistics prevail, $C(t)$ grows
more slowly and monotonically, saturating at a lower value without a pronounced
intermediate peak.

This consistent correspondence, observed across both local and non-local variants of the
mixed-field Ising model, underscores that the saturation value and dynamical profile of
Krylov complexity align robustly with spectral indicators of chaos. Thus, the joint use of
$C(t)$ and $\bar{r}$ provides a more nuanced and reliable characterization of the
integrable-to-chaotic transition than either measure alone.

In this work, we have systematically investigated signatures of quantum chaos in both local
and non-local variants of the mixed-field Ising spin chain, employing two complementary
diagnostics: the statistics of energy level spacings and the dynamical growth of Krylov
complexity. Our analysis demonstrates that non-local all-to-all interactions significantly
enhance the onset of chaos, with even weak couplings ($|g| > 0.25$) driving the spectral
statistics from Poisson to Wigner--Dyson distributions, in contrast to the local model, which
requires fine-tuning of field parameters to exhibit chaotic behavior.

The Krylov complexity $C(t)$ provides a real-time dynamical counterpart to these spectral
features: in chaotic regimes, it exhibits rapid growth, a characteristic peak, and saturates
at a higher plateau, whereas in integrable phases its evolution is slower, monotonic, and
saturates at lower values. The consistent correlation between the saturation value of $C(t)$
and the average level spacing ratio $\bar{r}$ underscores that these diagnostics together
offer a robust, multifaceted characterization of the integrable-to-chaotic transition.

Our results highlight the role of non-locality in accelerating the emergence of quantum
chaos and validate Krylov complexity as an effective dynamical probe, capable of
distinguishing chaotic from integrable dynamics in both local and long-range interacting
quantum spin chains.

\section{Conclusions}
Quantum chaos investigates how signatures of classical chaotic dynamics emerge in quantum systems. A central aspect is the manifestation of level repulsion in the energy spectrum, where energy eigenvalues become correlated and avoid degeneracies, reflecting underlying chaotic behavior. This correlation leads to characteristic Wigner-Dyson statistics for the distribution of level spacings, contrasting with the Poisson statistics typical of integrable systems. In the context of Ising spin chains, both local and non-local variants can display quantum chaotic features; non-local interactions often enhance complexity and thus amplify signatures of chaos.\\
The dynamics of operator spreading, as captured by Krylov complexity, further illuminate these distinctions. In chaotic regimes, Krylov complexity grows rapidly at early times, signifying the exponential sensitivity to initial conditions and the extensive delocalization of operators over the accessible Hilbert space. This growth frequently exhibits a pronounced peak before eventually saturating due to finite system size. In contrast, integrable systems generally show slower, more monotonic growth of Krylov complexity without such a distinct peak before saturation, reflecting their regular and predictable dynamics.\\
A widely used spectral diagnostic complementary to level spacing distributions is the average level spacing ratio, defined by: 
\begin{equation}\label{eq.28}
\bar{r}=\frac{1}{D_0-1}\sum_{n=1}^{D_0-1}min\left(r_j,\frac{1}{r_j}\right),\ \ r_j=\frac{\mathcal{S}_j}{\mathcal{S}_j+1}
\end{equation}\\
Where $\mathcal{S}_j=E_{j+1}-E_j$. For the Poisson model ${\bar{r}}_{poisson}\approx0.3863$ and for the Gaussian orthogonal matrix (GOE) ${\bar{r}}_{GOE}=0.5307$.\\
Our numerical results, illustrated in Fig. \ref{Fig.5}, demonstrate that for interaction strengths $| g |>0.25$, the non-local Ising model exhibits clear signatures of chaos, consistent with elevated level spacing ratios and enhanced Krylov complexity. In the local scenario, a fully chaotic phase emerges at $h_{z}=0.5$, while the system remains integrable at $h_{z}=0$. These observations are corroborated by the behavior of Krylov complexity, which aligns with the spectral diagnostics across these parameter regimes.\\

\section*{Acknowledgment}
We gratefully acknowledge Mohammad Reza Tanhayi for his guidance and support.  We wish to thank M. Alishahiha for his support. We would like to extend special thanks to M. Hosseinian, Z. Mousavi , Sh. Mirchi , F. Tayefi, and M. Keramat for their support. The work of Reza Pirmoradian
is based on research funded by the Iran National Science Foundation (INSF) under Project
No. 4026389. Lastly, we recognize the use of AI tools in assisting with text editing and
refinement.


\begin{thebibliography}{99}

\bibitem{Srednicki:1994mfb}
M.~Srednicki,
``Chaos and Quantum Thermalization,''
Phys. Rev. E \textbf{50}, 888
doi:10.1103/PhysRevE.50.888
[arXiv:cond-mat/9403051 [cond-mat]].

\bibitem{Deutsch1991}
J.~M.~Deutsch,
``Quantum statistical mechanics in a closed system,''
Phys. Rev. A \textbf{43}, no.4, 2046 (1991)
doi:10.1103/PhysRevA.43.2046.

\bibitem{Banuls:2010zki}
M.~C.~Ba{\~n}uls, J.~I.~Cirac and M.~B.~Hastings,
``Strong and Weak Thermalization of Infinite Nonintegrable Quantum Systems,''
Phys. Rev. Lett. \textbf{106}, no.5, 050405 (2011)
doi:10.1103/PhysRevLett.106.050405
[arXiv:1007.3957 [quant-ph]].

\bibitem{Hosur:2015ylk}
P.~Hosur, X.~L.~Qi, D.~A.~Roberts and B.~Yoshida,
``Chaos in quantum channels,''
JHEP \textbf{02}, 004 (2016)
doi:10.1007/JHEP02(2016)004
[arXiv:1511.04021 [hep-th]].

\bibitem{Maldacena:2015waa}
J.~Maldacena, S.~H.~Shenker and D.~Stanford,
``A bound on chaos,''
JHEP \textbf{08}, 106 (2016)
doi:10.1007/JHEP08(2016)106
[arXiv:1503.01409 [hep-th]].

\bibitem{Kitaev:2017awl}
A.~Kitaev and S.~J.~Suh,
``The soft mode in the Sachdev-Ye-Kitaev model and its gravity dual,''
JHEP \textbf{05}, 183 (2018)
doi:10.1007/JHEP05(2018)183
[arXiv:1711.08467 [hep-th]].

\bibitem{Roberts:2018mnp}
D.~A.~Roberts, D.~Stanford and A.~Streicher,
``Operator growth in the SYK model,''
JHEP \textbf{06}, 122 (2018)
doi:10.1007/JHEP06(2018)122
[arXiv:1802.02633 [hep-th]].

\bibitem{vonKeyserlingk:2017dyr}
C.~von Keyserlingk, T.~Rakovszky, F.~Pollmann and S.~Sondhi,
``Operator hydrodynamics, OTOCs, and entanglement growth in systems without conservation laws,''
Phys. Rev. X \textbf{8}, no.2, 021013 (2018)
doi:10.1103/PhysRevX.8.021013
[arXiv:1705.08910 [cond-mat.str-el]].

\bibitem{Nahum:2017yvy}
A.~Nahum, S.~Vijay and J.~Haah,
``Operator Spreading in Random Unitary Circuits,''
Phys. Rev. X \textbf{8}, no.2, 021014 (2018)
doi:10.1103/PhysRevX.8.021014
[arXiv:1705.08975 [cond-mat.str-el]].

\bibitem{Aleiner:2016eni}
I.~L.~Aleiner, L.~Faoro and L.~B.~Ioffe,
``Microscopic model of quantum butterfly effect: out-of-time-order correlators and traveling combustion waves,''
Annals Phys. \textbf{375}, 378-406 (2016)
doi:10.1016/j.aop.2016.09.006
[arXiv:1609.01251 [cond-mat.stat-mech]].

\bibitem{Chowdhury:2017jzb}
D.~Chowdhury and B.~Swingle,
``Onset of many-body chaos in the $O(N)$ model,''
Phys. Rev. D \textbf{96}, 065005 (2017)
doi:10.1103/PhysRevD.96.065005
[arXiv:1703.02545 [cond-mat.str-el]].

\bibitem{Garttner:2016mqj}
M.~G{\"a}rttner, J.~G.~Bohnet, A.~Safavi-Naini, M.~L.~Wall, J.~J.~Bollinger and A.~M.~Rey,
``Measuring out-of-time-order correlations and multiple quantum spectra in a trapped ion quantum magnet,''
Nature Phys. \textbf{13}, 781 (2017)
doi:10.1038/nphys4119
[arXiv:1608.08938 [quant-ph]].

\bibitem{Li:2016xhw}
J.~Li, R.~Fan, H.~Wang, B.~Ye, B.~Zeng, H.~Zhai, X.~Peng and J.~Du,
``Measuring Out-of-Time-Order Correlators on a Nuclear Magnetic Resonance Quantum Simulator,''
Phys. Rev. X \textbf{7}, no.3, 031011 (2017)
doi:10.1103/PhysRevX.7.031011
[arXiv:1609.01246 [cond-mat.str-el]].

\bibitem{Lin2018}
C.-J.~Lin and O.~I.~Motrunich,
``Out-of-time-ordered correlators in a quantum Ising chain,''
Phys. Rev. B \textbf{97}, no.14, 144304 (2018)
doi:10.1103/PhysRevB.97.144304
[arXiv:1801.01636 [cond-mat.stat-mech]].

\bibitem{Lin2018b}
C.-J.~Lin and O.~I.~Motrunich,
``Out-of-time-ordered correlators in short-range and long-range hard-core boson models and in the Luttinger-liquid model,''
Phys. Rev. B \textbf{98}, no.13, 134305 (2018)
doi:10.1103/PhysRevB.98.134305
[arXiv:1807.08826 [cond-mat.str-el]].

\bibitem{Gopalakrishnan:2018rfu}
S.~Gopalakrishnan,
``Operator growth and eigenstate entanglement in an interacting integrable Floquet system,''
Phys. Rev. B \textbf{98}, no.6, 060302 (2018)
doi:10.1103/PhysRevB.98.060302
[arXiv:1806.04156 [cond-mat.stat-mech]].

\bibitem{Khemani:2018sdn}
V.~Khemani, D.~A.~Huse and A.~Nahum,
``Velocity-dependent Lyapunov exponents in many-body quantum, semiclassical, and classical chaos,''
Phys. Rev. B \textbf{98}, no.14, 144304 (2018)
doi:10.1103/PhysRevB.98.144304
[arXiv:1803.05902 [cond-mat.stat-mech]].

\bibitem{Xu:2018xfz}
S.~Xu and B.~Swingle,
``Accessing scrambling using matrix product operators,''
Nature Phys. \textbf{16}, no.2, 199-204 (2019)
doi:10.1038/s41567-019-0712-4
[arXiv:1802.00801 [quant-ph]].

\bibitem{Sunderhauf:2019djv}
C.~S{\"u}nderhauf, L.~Piroli, X.~L.~Qi, N.~Schuch and J.~I.~Cirac,
``Quantum chaos in the Brownian SYK model with large finite $N$: OTOCs and tripartite information,''
JHEP \textbf{11}, 038 (2019)
doi:10.1007/JHEP11(2019)038
[arXiv:1908.00775 [quant-ph]].

\bibitem{Yan:2019fbg}
B.~Yan, L.~Cincio and W.~H.~Zurek,
``Information Scrambling and Loschmidt Echo,''
Phys. Rev. Lett. \textbf{124}, no.16, 160603 (2020)
doi:10.1103/PhysRevLett.124.160603
[arXiv:1903.02651 [quant-ph]].

\bibitem{Rozenbaum:2016mmv}
E.~B.~Rozenbaum, S.~Ganeshan and V.~Galitski,
``Lyapunov Exponent and Out-of-Time-Ordered Correlator{\textquoteright}s Growth Rate in a Chaotic System,''
Phys. Rev. Lett. \textbf{118}, no.8, 086801 (2017)
doi:10.1103/PhysRevLett.118.086801
[arXiv:1609.01707 [cond-mat.dis-nn]].

\bibitem{Sachdev1993}
S.~Sachdev and J.~Ye,
``Gapless spin-fluid ground state in a random quantum Heisenberg magnet,''
Phys. Rev. Lett. \textbf{70}, 3339 (1993)
doi:10.1103/PhysRevLett.70.3339
[arXiv:cond-mat/9212030].

\bibitem{Maldacena:2016hyu}
J.~Maldacena and D.~Stanford,
``Remarks on the Sachdev-Ye-Kitaev model,''
Phys. Rev. D \textbf{94}, no.10, 106002 (2016)
doi:10.1103/PhysRevD.94.106002
[arXiv:1604.07818 [hep-th]].

\bibitem{Fine:2013zog}
B.~V.~Fine, T.~A.~Elsayed, C.~M.~Kropf and A.~S.~de Wijn,
``Absence of exponential sensitivity to small perturbations in nonintegrable systems of spins 1/2,''
Phys. Rev. E \textbf{89}, 012923 (2014)
doi:10.1103/PhysRevE.89.012923
[arXiv:1305.2817 [cond-mat.stat-mech]].

\bibitem{Hashimoto:2017oit}
K.~Hashimoto, K.~Murata and R.~Yoshii,
``Out-of-time-order correlators in quantum mechanics,''
JHEP \textbf{10}, 138 (2017)
doi:10.1007/JHEP10(2017)138
[arXiv:1703.09435 [hep-th]].

\bibitem{Hashimoto:2020xfr}
K.~Hashimoto, K.~B.~Huh, K.~Y.~Kim and R.~Watanabe,
``Exponential growth of out-of-time-order correlator without chaos: inverted harmonic oscillator,''
JHEP \textbf{11}, 068 (2020)
doi:10.1007/JHEP11(2020)068
[arXiv:2007.04746 [hep-th]].

\bibitem{Doroudiani:2019llj}
M.~Doroudiani, A.~Naseh and R.~Pirmoradian,
``Complexity for Charged Thermofield Double States,''
JHEP \textbf{01}, 120 (2020)
doi:10.1007/JHEP01(2020)120
[arXiv:1910.08806 [hep-th]].

\bibitem{Pirmoradian2020}
R.~Pirmoradian and M.~R.~Tanhayi,
``On the Complexity of a Charged Quantum Oscillator,''
J. Korean Phys. Soc. \textbf{77}, no.2, 102 (2020)
doi:10.3938/jkps.77.102
[arXiv:1911.08886 [physics.gen-ph]].

\bibitem{RezaTanhayi:2018cyv}
M.~Reza Tanhayi, R.~Vazirian and S.~Khoeini-Moghaddam,
``Complexity Growth Following Multiple Shocks,''
Phys. Lett. B \textbf{790}, 49-57 (2019)
doi:10.1016/j.physletb.2018.12.067
[arXiv:1809.05044 [hep-th]].

\bibitem{Khorasani:2023usq}
F.~Khorasani, R.~Pirmoradian and M.~R.~Tanhayi,
``Position dependence of Nielsen complexity for the thermofield double state,''
Phys. Lett. B \textbf{851}, 138585 (2024)
doi:10.1016/j.physletb.2024.138585
[arXiv:2308.15836 [quant-ph]].

\bibitem{Pirmoradian2025}
R. Pirmoradian, N. Abolghasemiazad, E. Sadoogh, Z. MohammadAli, M. Teymouri, “Defining Fundamental Computational Speed Limits for Quantum Oscillators,” Int. J. Math. Model. Comput. 15, 4 (2025)
doi:10.57647/ijm2c.2025.15041.

\bibitem{Ghasemi:2021jiy}
M.~Ghasemi, A.~Naseh and R.~Pirmoradian,
``Odd entanglement entropy and logarithmic negativity for thermofield double states,''
JHEP \textbf{10}, 128 (2021)
doi:10.1007/JHEP10(2021)128
[arXiv:2106.15451 [hep-th]].

\bibitem{Pirmoradian:2023uvt}
R.~Pirmoradian and M.~R.~Tanhayi,
``Symmetry-resolved entanglement entropy for local and non-local QFTs,''
Eur. Phys. J. C \textbf{84}, no.8, 849 (2024)
doi:10.1140/epjc/s10052-024-13212-8
[arXiv:2311.00494 [hep-th]].

\bibitem{Pirmoradian:2025nxw}
R.~Pirmoradian, M.~H.~Bek-Khoshnevis, S.~Ebadi and M.~R.~Tanhayi,
``Entanglement Structure of Nonlocal Field Theories,''
[arXiv:2511.10505 [quant-ph]].

\bibitem{Pirmoradian:2021wvo}
R.~Pirmoradian and M.~R.~Tanhayi,
``Non-local probes of entanglement in the scale-invariant gravity,''
Int. J. Geom. Meth. Mod. Phys. \textbf{18}, no.12, 2150197 (2021)
doi:10.1142/S0219887821501978
[arXiv:2103.02998 [hep-th]].

\bibitem{Pirmoradian:2025dco}
R.~Pirmoradian and M.~R.~Tanhayi,
``Information Dynamics in Quantum Harmonic Systems: Insights from Toy Models,''
[arXiv:2501.14359 [quant-ph]].

\bibitem{Prosen:2007gfp}
T.~Prosen and I.~Pi{\v{z}}orn,
``Operator space entanglement entropy in a transverse Ising chain,''
Phys. Rev. A \textbf{76}, no.3, 032316 (2007)
doi:10.1103/PhysRevA.76.032316
[arXiv:0706.2480 [quant-ph]].

\bibitem{Pizorn:2009gup}
I.~Pi{\v{z}}orn and T.~Prosen,
``Operator space entanglement entropy in XY spin chains,''
Phys. Rev. B \textbf{79}, no.18, 184416 (2009)
doi:10.1103/PhysRevB.79.184416
[arXiv:0903.2432 [quant-ph]].

\bibitem{Alba:2019okd}
V.~Alba, J.~Dubail and M.~Medenjak,
``Operator Entanglement in Interacting Integrable Quantum Systems: The Case of the Rule 54 Chain,''
Phys. Rev. Lett. \textbf{122}, no.25, 250603 (2019)
doi:10.1103/PhysRevLett.122.250603
[arXiv:1901.04521 [cond-mat.stat-mech]].

\bibitem{Bertini:2019gbu}
B.~Bertini, P.~Kos and T.~Prosen,
``Operator Entanglement in Local Quantum Circuits I: Chaotic Dual-Unitary Circuits,''
SciPost Phys. \textbf{8}, no.4, 067 (2020)
doi:10.21468/SciPostPhys.8.4.067
[arXiv:1909.07407 [cond-mat.stat-mech]].

\bibitem{MacCormack:2020auw}
I.~MacCormack, M.~T.~Tan, J.~Kudler-Flam and S.~Ryu,
``Operator and entanglement growth in nonthermalizing systems: Many-body localization and the random singlet phase,''
Phys. Rev. B \textbf{104}, no.21, 214202 (2021)
doi:10.1103/PhysRevB.104.214202
[arXiv:2001.08222 [cond-mat.str-el]].

\bibitem{Mascot:2020qep}
E.~Mascot, M.~Nozaki and M.~Tezuka,
``Local operator entanglement in spin chains,''
SciPost Phys. Core \textbf{6}, 070 (2023)
doi:10.21468/SciPostPhysCore.6.4.070
[arXiv:2012.14609 [cond-mat.dis-nn]].

\bibitem{Alba:2020npy}
V.~Alba,
``Diffusion and operator entanglement spreading,''
Phys. Rev. B \textbf{104}, no.9, 094410 (2021)
doi:10.1103/PhysRevB.104.094410
[arXiv:2006.02788 [cond-mat.stat-mech]].

\bibitem{Parker:2018yvk}
D.~E.~Parker, X.~Cao, A.~Avdoshkin, T.~Scaffidi and E.~Altman,
``A Universal Operator Growth Hypothesis,''
Phys. Rev. X \textbf{9}, no.4, 041017 (2019)
doi:10.1103/PhysRevX.9.041017
[arXiv:1812.08657 [cond-mat.stat-mech]].

\bibitem{Barbon:2019wsy}
J.~L.~F.~Barb{\'o}n, E.~Rabinovici, R.~Shir and R.~Sinha,
``On The Evolution Of Operator Complexity Beyond Scrambling,''
JHEP \textbf{10}, 264 (2019)
doi:10.1007/JHEP10(2019)264
[arXiv:1907.05393 [hep-th]].

\bibitem{Vasli:2023syq}
M.~J.~Vasli, K.~Babaei Velni, M.~R.~Mohammadi Mozaffar, A.~Mollabashi and M.~Alishahiha,
``Krylov complexity in Lifshitz-type scalar field theories,''
Eur. Phys. J. C \textbf{84}, no.3, 235 (2024)
doi:10.1140/epjc/s10052-024-12609-9
[arXiv:2307.08307 [hep-th]].

\bibitem{Viswanath:1994}
V.~S.~Viswanath and G.~Mueller,  
``The recursion method: Application to many-body dynamics,''  
Springer, \textbf{23}, (1994).  
doi:10.1007/978-3-540-48651-0

\bibitem{Dymarsky:2021bjq}
A.~Dymarsky and M.~Smolkin,
``Krylov complexity in conformal field theory,''
Phys. Rev. D \textbf{104}, no.8, L081702 (2021)
doi:10.1103/PhysRevD.104.L081702
[arXiv:2104.09514 [hep-th]].

\bibitem{Bhattacharjee:2022vlt}
B.~Bhattacharjee, X.~Cao, P.~Nandy and T.~Pathak,
``Krylov complexity in saddle-dominated scrambling,''
JHEP \textbf{05}, 174 (2022)
doi:10.1007/JHEP05(2022)174
[arXiv:2203.03534 [quant-ph]].

\bibitem{Avdoshkin:2022xuw}
A.~Avdoshkin, A.~Dymarsky and M.~Smolkin,
``Krylov complexity in quantum field theory, and beyond,''
JHEP \textbf{06}, 066 (2024)
doi:10.1007/JHEP06(2024)066
[arXiv:2212.14429 [hep-th]].

\bibitem{Camargo:2022rnt}
H.~A.~Camargo, V.~Jahnke, K.~Y.~Kim and M.~Nishida,
``Krylov complexity in free and interacting scalar field theories with bounded power spectrum,''
JHEP \textbf{05}, 226 (2023)
doi:10.1007/JHEP05(2023)226
[arXiv:2212.14702 [hep-th]].

\bibitem{Rabinovici:2021qqt}
E.~Rabinovici, A.~S{\'a}nchez-Garrido, R.~Shir and J.~Sonner,
``Krylov localization and suppression of complexity,''
JHEP \textbf{03}, 211 (2022)
doi:10.1007/JHEP03(2022)211
[arXiv:2112.12128 [hep-th]].

\bibitem{Rabinovici:2022beu}
E.~Rabinovici, A.~S{\'a}nchez-Garrido, R.~Shir and J.~Sonner,
``Krylov complexity from integrability to chaos,''
JHEP \textbf{07}, 151 (2022)
doi:10.1007/JHEP07(2022)151
[arXiv:2207.07701 [hep-th]].

\bibitem{Alishahiha:2022nhe}
M.~Alishahiha,
``On quantum complexity,''
Phys. Lett. B \textbf{842}, 137979 (2023)
doi:10.1016/j.physletb.2023.137979
[arXiv:2209.14689 [hep-th]].

\bibitem{Baggioli:2024wbz}
M.~Baggioli, K.~B.~Huh, H.~S.~Jeong, K.~Y.~Kim and J.~F.~Pedraza,
``Krylov complexity as an order parameter for quantum chaotic-integrable transitions,''
Phys. Rev. Res. \textbf{7}, no.2, 023028 (2025)
doi:10.1103/PhysRevResearch.7.023028
[arXiv:2407.17054 [hep-th]].

\bibitem{Bohigas:1983er}
O.~Bohigas, M.~J.~Giannoni and C.~Schmit,
``Characterization of chaotic quantum spectra and universality of level fluctuation laws,''
Phys. Rev. Lett. \textbf{52}, 1-4 (1984)
doi:10.1103/PhysRevLett.52.1.

\bibitem{DAlessio:2015qtq}
L.~D'Alessio, Y.~Kafri, A.~Polkovnikov and M.~Rigol,
``From quantum chaos and eigenstate thermalization to statistical mechanics and thermodynamics,''
Adv. Phys. \textbf{65}, no.3, 239-362 (2016)
doi:10.1080/00018732.2016.1198134
[arXiv:1509.06411 [cond-mat.stat-mech]].

\bibitem{Chan:2018dzt}
A.~Chan, A.~De Luca and J.~T.~Chalker,
``Spectral statistics in spatially extended chaotic quantum many-body systems,''
Phys. Rev. Lett. \textbf{121}, no.6, 060601 (2018)
doi:10.1103/PhysRevLett.121.060601
[arXiv:1803.03841 [cond-mat.stat-mech]].

\bibitem{Kawabata:2021tsf}
K.~Kawabata and S.~Ryu,
``Nonunitary Scaling Theory of Non-Hermitian Localization,''
Phys. Rev. Lett. \textbf{126}, 166801 (2021)
doi:10.1103/PhysRevLett.126.166801
[arXiv:2005.00604 [cond-mat.dis-nn]].

\bibitem{Muck:2022xfc}
W.~M{\"u}ck and Y.~Yang,
``Krylov complexity and orthogonal polynomials,''
Nucl. Phys. B \textbf{984}, 115948 (2022)
doi:10.1016/j.nuclphysb.2022.115948
[arXiv:2205.12815 [hep-th]].

\bibitem{Craps:2019rbj}
B.~Craps, M.~De Clerck, D.~Janssens, V.~Luyten and C.~Rabideau,
``Lyapunov growth in quantum spin chains,''
Phys. Rev. B \textbf{101}, no.17, 174313 (2020)
doi:10.1103/PhysRevB.101.174313
[arXiv:1908.08059 [hep-th]].

\bibitem{Caceres:2024mec}
E.~C{\'a}ceres, S.~Eccles, J.~Pollack and S.~Racz,
``Generic ETH: Eigenstate Thermalization beyond the Microcanonical,''
[arXiv:2403.05197 [quant-ph]].

\bibitem{Belyansky:2020bia}
R.~Belyansky, P.~Bienias, Y.~A.~Kharkov, A.~V.~Gorshkov and B.~Swingle,
``Minimal Model for Fast Scrambling,''
Phys. Rev. Lett. \textbf{125}, no.13, 130601 (2020)
doi:10.1103/PhysRevLett.125.130601
[arXiv:2005.05362 [quant-ph]].

\end{thebibliography}
\end{document}